\documentstyle[eqsecnum,aps,amstex,epsfig,float,amssymb]{revtex}

\begin{document}

\newcommand{\light}{\operatorname{c}}
\newcommand{\e}{\operatorname{e}}

\newcommand{\aeta}{{\em Astron. Astrophys.}}
\newcommand{\astrophysj}{{\em Astrophys. Jour.}}
\newcommand{\mnras}{{\em Mon. Not. R. Ast. Soc.}}
\newcommand{\nuclphysa}{{\em Nucl. Phys. A}}
\newcommand{\physrevd}{{\em Phys. Rev. D}}
\newcommand{\physrevl}{{\em Phys. Rev. Lett.}}
\newcommand{\apjlet}{{\em Astrophys. Jour. Lett.}}
\newcommand{\cqg}{{\em Class. Quantum Grav.}}

\draft

\title{Coincidence and coherent data analysis methods for gravitational wave bursts in a network of interferometric detectors}

\author{Nicolas Arnaud, Matteo Barsuglia, Marie-Anne Bizouard, Violette Brisson, Fabien Cavalier,\\Michel Davier, Patrice Hello, Stephane Kreckelbergh and Edward K. Porter}

\address{Laboratoire de l'Acc\'el\'erateur Lin\'eaire, IN2P3-CNRS and Universit\'e Paris Sud,\\B.P. 34, B\^atiment 200, Campus d'Orsay, 91898 Orsay Cedex (France)\protect\\}

\maketitle

\begin{abstract}

Network data analysis methods are the only way to properly separate real gravitational wave (GW) transient events from detector noise. They can be divided into two generic classes: the coincidence method and the coherent analysis. The former uses lists of selected events provided by each interferometer belonging to the network and tries to correlate them in time to identify a physical signal. Instead of this binary treatment of detector outputs (signal present or absent), the latter method involves first the merging of the interferometer data and looks for a common pattern, consistent with an assumed GW waveform and a given source location in the sky. The thresholds are only applied later, to validate or not the hypothesis made. As coherent algorithms use a more complete information than coincidence methods, they are expected to provide better detection performances, but at a higher computational cost. An efficient filter must yield a good compromise between a low false alarm rate (hence triggering on data at a manageable rate) and a high detection efficiency. Therefore, the comparison of the two approaches is achieved using so-called Receiving Operating Characteristics (ROC), giving the relationship between the false alarm rate and the detection efficiency for a given method. This paper investigates this question via Monte-Carlo simulations, using the network model developed in a previous article. Its main conclusions are the following. First, a three-interferometer network like Virgo-LIGO is found to be too small to reach good detection efficiencies at low false alarm rates: larger configurations are suitable to reach a confidence level high enough to validate as true GW a detected event. In addition, an efficient network must contain interferometers with comparable sensitivities: studying the three-interferometer LIGO network shows that the 2-km interferometer with half sensitivity leads to a strong reduction of performances as compared to a network of three interferometers with full sensitivity. Finally, it is shown that coherent analyses are feasible for burst searches and are clearly more efficient than coincidence strategies. Therefore, developing such methods should be an important goal of a worldwide collaborative data analysis.

\end{abstract}

\pacs{PACS numbers 04.80.Nn, 07.05.Kf}


\section{Introduction}

The first generation of large interferometric gravitational wave (GW) detectors \cite{geo,ligo,tama,virgo,aciga} is producing a growing set of experimental results, showing that the detectors already built are coming close to their foreseen sensitivities. Therefore, it is very important to consider exchanging data and perform analysis in common. Such network data analysis methods are compulsory to separate -- with a sufficient confidence level -- real GW transient signals from noise occurring in one particular detector.

GW burst signals have usually a small duration (a few ms) and a poorly known shape: for instance, type II supernovae or the merging phase of coalescing compact binary systems belong to this category. As they are not accurately modeled, only suboptimal methods \cite{arnaud_burst_1,bala,mohanty,powermonit,pradier,vicere} can be used to detect them. Therefore, various related outputs coming from a set of interferometers are required to reach a definite conclusion on the reality of the GW event. In addition to the definition of efficient filters suitable to analyze single interferometer outputs, it is also necessary to estimate the performances of different network data analysis methods. Studying this problem with Monte-Carlo simulations is the goal of this paper.

Network data analysis methods can be classified into two categories: coincidence or coherent filtering. The first method is simpler to use and has been already considered for years. In this approach, each interferometer belonging to the network analyzes separately its own data and produces a list of selected events, characterized by their timing and their maximum signal-to-noise ratio (SNR) exceeding some suitable threshold levels. In a second step, these events are correlated with those found by other detectors in order to see if some are compatible with a real GW source. Many articles in the literature \cite{gursel_tinto,jk5,arnaud_burst_2} deal with this topic, in particular with real data taken by resonant bar experiments \cite{IGEC,IGEC_papers}.

As pointed out independently \cite{Bose,Finn}, such coincidence analyzes are not optimal in the sense that their binary use of interferometer data (a GW signal in a given detector is either present or absent) leaves aside important information on the possible correlations between the different datasets. Indeed, as the detector beam patterns are not uniform \cite{thorne87}, the interaction between the physical signal and a given interferometer depends on the relative location of the source. Thus, the GW amplitude scales differently in the various components of the network causing, in addition to noise fluctuations, the filtered outputs to exceed or not the selection threshold. These large variations of the GW response could suppress even a strong GW signal in a given interferometer\cite{arnaud_burst_2}. In a coherent analysis, all detector outputs contribute to the filtering algorithm. Triggered and merged in a suitable way, a higher statistical significance can be achieved compared to individual analyzes, thus improving the network detection potential.

Coherent methods have already been studied in the literature, particularly for the search of signals with known waveforms. Both Ref. \cite{Bose} and \cite{Finn} consider network data analysis methods based on a likelihood function, which turn out to be direct extensions of the Wiener filtering, optimal for a single detector search when the signal shape is known. Coincidence and coherent detections have been compared in a two-detector network \cite{Finn}, with a simple model describing its interaction with a GW burst signal (two sinusoid cycles). The coherent search was found to be always better than the coincidence method and the results are robust with respect to the noise statistics. Ref. \cite{Bose}, later extended in \cite{PDB}, deals with in-spiral waveforms at Newtonian order. A new formalism is developed based on a likelihood function, in a way very similar to matched filtering, but now with a parameter space containing two more unknowns which correspond to the source position in the sky: consequently, the data analysis procedure becomes more computationally expensive. Ref. \cite{Bose2,Bose3} applied then this framework to post-Newtonian inspiral waveforms and to the case of a non-Gaussian noise.

The present study takes advantage of these pioneering works and applies the same kind of methods for GW bursts, using the network detection model originally developed in \cite{arnaud_burst_2} to study coincidences between interferometers. On the one hand, this framework is simple enough to perform a large number of Monte-Carlo simulations, and thus to compare accurately coincident and coherent detections. On the other hand, all the features characterizing the interaction between a GW and a network of interferometers are properly taken into account: non uniform angular pattern of the detectors, propagation time delays between them and data sampling.

As an efficient detection algorithm must be a good compromise between a high detection probability and a low false alarm rate, a standard tool to estimate the performances of a data analysis method is to use ROC ({\it Receiver Operating Characteristics}). Such diagrams present the detection probability of a given signal (scaled at a particular SNR) versus the false alarm rate.

Section \ref{section:hyp_notations} summarizes the general framework of this study, specifying the network detection model used in all the simulations presented in this paper, and the tools used. The single detector performances -- detection efficiency and timing resolution -- are briefly recalled in Section \ref{section:single} as they are the basis for the following investigations. Then, the network coincidence analysis is studied in Section \ref{section:coincidences}. ROC studies for networks from two to six detectors -- comprising the large interferometers currently being developed, GEO600 \cite{geo}, the two 4-km LIGO detectors and the 2-km interferometer located in Hanford \cite{ligo}, TAMA300 \cite{tama}, Virgo \cite{virgo} and finally the foreseen ACIGA project \cite{aciga} -- are presented. To decide whether or not a coincidence is valid, two different compatibility conditions are considered. The first one, called 'loose' condition, does not require the knowledge of the source position in the sky while the second one, the 'tight' condition, does.

Section \ref{section:coherent_analysis} deals with the coherent analysis. First, the derivation of the likelihood statistics follows closely the analysis of Ref. \cite{PDB}; then, the corresponding ROC for coherent search of GW bursts are presented. As coherent filtering requires the knowledge of the source location in the sky, a set of coherent filters must be used in parallel to cover the full celestial sphere. Therefore, Section \ref{section:sky_tiling} estimates the number $\mathfrak{N}$ of such templates needed for a complete tiling of the sky, given a prescription on the maximal loss of SNR allowed. Knowing $\mathfrak{N}$ allows one to tabulate the false alarm rate of the coherent ROC. Finally, Section \ref{section:comparison} compares the two network data analysis methods considered in this article.

\section{Hypothesis and notations}
\label{section:hyp_notations}

\subsection{Interferometer response to a GW}
\label{subsection:ITF_GW}

Measuring the strength of a signal with respect to the background noise is not the only information needed to estimate how well a GW may be detected in an interferometric detector. As the angular pattern of an antenna is not uniform, it is also necessary to take into account the location of the source in the sky. The result $h(t)$ of the interaction between the wave and the instrument is a linear combination of the two GW polarizations $h_+$ and $h_\times$ \cite{thorne87}:

\begin{equation}
h(t) \; = \; F_+(t) \, h_+(t) \; + \; F_\times(t) \, h_\times(t)
\label{eq:response}
\end{equation}

The two weighting factors $F_+$ and $F_\times$ are called beam pattern functions whose values are between -1 and 1. They depend on many parameters which may be roughly classified into three sets.

(S1) The detector coordinates (longitude, latitude, orientation with respect to the local North-South direction).

(S2) The source location in the sky, given for instance by the celestial sphere coordinates (the right ascension $\alpha$ and the declination $\delta$) and the local sidereal time which takes into account the Earth proper rotational motion.

(S3) A vector of physical parameters describing the time evolution of the GW signal. Some may be estimated at the output of the data analysis procedure, but this always requires adding some hypothesis on the signal. The polarization angle $\psi$ is also included in this set.

In fact, the signal dependence on $\psi$ can be explicitly extracted. For an interferometer, one has:
\begin{equation}
\begin{pmatrix}
F_+(t) \\ F_\times(t)
\end{pmatrix}
=
\sin\chi \;
\begin{pmatrix}
\cos2\psi & \sin2\psi \\
-\sin2\psi &\cos2\psi
\end{pmatrix}
\begin{pmatrix}
a(t) \\ b(t)
\end{pmatrix}
\label{eq:BP}
\end{equation}
where $a(t)$ and $b(t)$ only depend on the sets (S1) and (S2) \cite{jk5,jks}, and with $\chi$ being the angle between the two arms of the interferometer.

\subsection{Interferometer network and source modeling}

Monte-Carlo simulations \cite{arnaud_burst_2} are used to compare the detection performances of coincident and coherent data analysis methods in various network configurations. As this paper aims at studying the consequences of the interferometer location on Earth rather than the effect of the current or foreseen differences in their sensitivities to GW, the network model uses the simplifying assumption of identical detection performances. In this way, all interferometers contribute equally to the network.

Yet, in Section \ref{subsection:LIGO_network}, a difference in the interferometer sensitivities is introduced in a complementary study, investigating the case of non-identical detectors. The 'LIGO network' (made of three interferometers, the two 4-km in Hanford and Livingston, and the 2-km in Hanford) is well-suited to this work: as the GW sensitivity should scale with the arm length, the 2-km detector should ultimately be half as sensitive as the two other LIGO interferometers. As this computation shows an important loss of efficiency induced by this difference in sensitivity, the Hanford 2-km detector is not considered elsewhere in this article.

The (S1) parameters of the detectors are chosen to match the already existing or planned instruments; as the local orientation of ACIGA is not yet defined, it has been optimized to maximize the detection efficiency in the full network of interferometers \cite{arnaud_burst_2,Searle}. The $P$ interferometers, labeled in the following by the index $i$, are assumed to have many features in common: the interaction with a GW signal -- as defined in Section \ref{subsection:ITF_GW} above --, the sampling frequency $f_{\text{samp}}$, and the noise characteristics. All noises are taken to be Gaussian, white and uncorrelated with the same RMS: $\sigma^i=\sigma$. Finally, the interferometers are assumed to be properly synchronized.

Any correlation between a filtering function $s(t)$ and the output $x^i(t)$ of the $i$-th interferometer at time $t$ (the sampling time at the origin of the analyzing windows) is represented in the time domain by the following quantity:

\begin{equation}
\langle \, s \, | \, x^i \, \rangle \, \left( t \right) \; = \; \sum_{k=0}^{{\mathbb{N}}-1} \; s\left( \frac{ k }{ f_{\text{samp}} } \right) \, \times \, x^i\left( t \, + \frac{ k }{ f_{\text{samp}} } \right)
\label{eq:filtering}
\end{equation}
with $\mathbb{N}$ being the filtering window size.

The GW signal $s(t)$ is assumed to be a Gaussian peak of 'half-width' $\omega=1$~ms: $s(t) \propto \exp(-t^2/2\omega^2)$. Such pulse-like shapes are characteristic of the most common GW burst waveforms simulated numerically: see e.g.\cite{ZM,DFM} for the case of supernova signals. Its amplitude with respect to the background noise is monitored by its optimal SNR $\rho_{\text{max}}$ \cite{arnaud_burst_2}: the average value of the filter output, computed in a noisy background with both the Wiener filtering method and a detector optimally oriented. In the following, we mainly focus on 'weak' signals for which detection problems are likely. We also assume that the sources are uniformly distributed over the sky, with a random timing. Finally, matched filtering is used to simulate the detection process.

\subsection{Receiving Operator Characteristics (ROC)}

A convenient way to estimate the capability of a given filter to detect a particular GW signal -- whose amplitude is fixed according to a chosen value of the  maximum SNR $\rho_{\text{max}}$ -- is the ROC, which presents on a single plot the detection efficiency $\epsilon$ versus the filter false alarm rate $\tau$ obtained by varying the threshold $\eta$. Following the prescription of Ref. \cite{Finn,hello_timing}, we consider in the rest of this paper a false alarm rate {\it per bin}: 

\begin{equation}
\tau_{\text{norm}} \; = \; \frac{ N_{\text{FA}} }{ {\mathbb{N}} \, \times N_{\text{MC}} }
\end{equation}
with $N_{\text{FA}}$ being the total number of false alarms, $N_{\text{MC}}$ the number of Monte-Carlo simulations, and ${\mathbb{N}}$ the size of the vector of data analyzed at each simulation loop.

\subsection{Coincidence analysis}
\label{subsection:coincidence_analysis}

For the coincidence analysis framework presented in this paper, the event compatibility is tested by comparing the delays between the triggered events for every pair of detectors. In case of a real GW signal (detected in two detectors $D^i$ and $D^j$ at times $t^i$ and $t^j$ respectively), the time difference is related to the source position in the sky. Let $\vec{n}$ be the unit vector pointing from the Earth center to the GW source location. One has:

\begin{equation}
\Delta t^{ij}(\vec{n}) \; = \; t^j \, - \, t^i \; = \; \frac{ \vec{n} \, \cdot \, \overrightarrow{D^jD^i} }{ c }
\label{eq:delay}
\end{equation}
if the filters have triggered on the GW signal in both detectors. Neglecting timing errors, such equation defines a circle in the sky on which the source is located.

Two compatibility tests can be set from Eq. (\ref{eq:delay}) for coincidence analysis: the first one -- the 'loose' test -- does not assume that the source location in the sky is known while the other -- the 'tight' test uses this additional information. The former case allows one to survey the whole sky with one single analysis, but at the price of a lower efficiency, whereas the latter can reject more false coincidence events with a more stringent compatibility condition.

\subsubsection{Loose compatibility}

Let us first consider a full sky search without any knowledge on the GW source location in the sky. The timing delay $\Delta t^{ij}$ between detectors $D^i$ and $D^j$ must obey the following inequality:

\begin{equation}
\Delta t^{ij} \; \lesssim \; \frac{ \| \overrightarrow{D^iD^j}\| }{ c } \; = \; \Delta t^{ij}_{\text{max}}
\label{eq:loose}
\end{equation}
with the term on the right side of the inequality being the light time travel between the two detectors. Table~I shows the maximum delays between all pairs of interferometers. The largest distance is between LIGO Livingston and ACIGA, about 42~ms.

\begin{center}
\begin{tabular}{|cccccc|}
\hline & LIGO Hanford & LIGO Livingston & GEO600 & TAMA300 & ACIGA \\
Virgo & 27.20 & 26.39 & 3.20 & 29.56 & 37.06 \\
LIGO Hanford & & 10.00 & 25.01 & 24.86 & 39.33 \\
LIGO Livingston & & & 25.04 & 32.24 & 41.68 \\
GEO600 & & & & 27.80 & 37.46 \\
TAMA300 & & & & & 24.58\\
\hline
\end{tabular}
\end{center}
\vskip -0.2truecm
\centerline{Table I: Maximum time delays (in ms) $\Delta t^{ij}_{\text{max}}$ between pairs of interferometers.}

\begin{figure}
\centerline{\epsfig{file=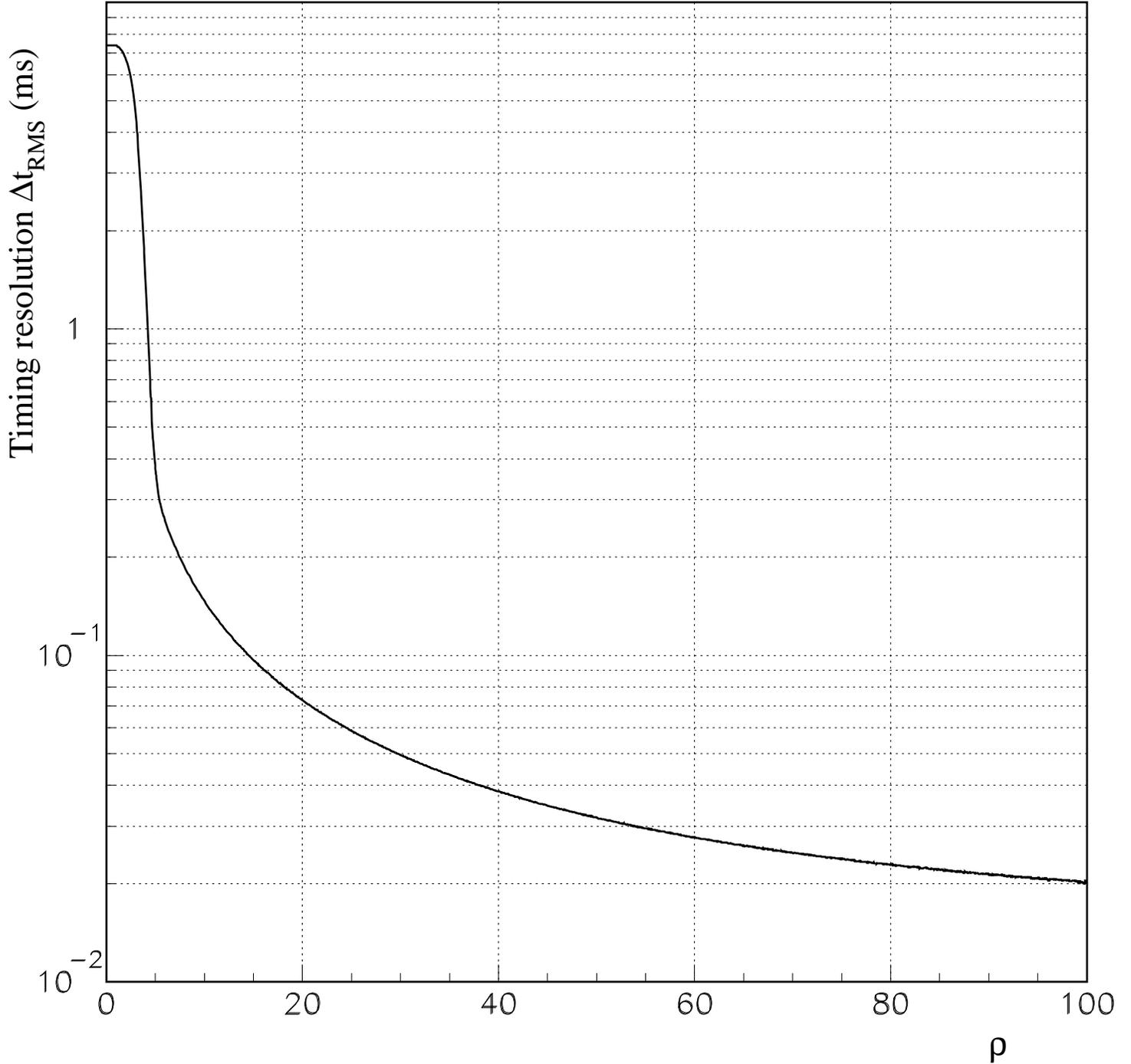,width=20cm}}
\caption{Timing resolution $\Delta t_{\text{RMS}}$ (in ms) of the Gaussian filter versus the SNR $\rho$ of the signal as detected in the interferometer. When $\rho \rightarrow 0$, the noise becomes dominant: the timing of the maximum output is uniformly distributed in the analysis window $({\mathbb{N}}=512)$ and the resolution reaches the plateau ${\mathbb{N}}/\sqrt{12}/f_{\text{sampling}}\approx 7.4$ ms. The analytical fit presented in [14] $\Delta t_{\text{RMS}}=1.45/\rho$ is valid in the intermediate range $\rho \in [6;30]$. For smaller values of the SNR, it underestimates the timing uncertainty which increases much faster because the noise contribution becomes more dominant. On the other hand, the fit overestimates $\Delta t_{\text{RMS}}$ for very high values of $\rho$.}
\label{fig:timing_resolution}
\end{figure}

To take into account the statistical uncertainty on the timing locations, an error must be associated with the delay $\Delta t^{ij}$. For each interferometer, a single detection error is computed by using the relation between $\Delta t_{\text{RMS}}$ and the maximum filter output $\rho$ shown on Figure \ref{fig:timing_resolution} -- see Section \ref{subsection:timing_performances} for more details. The two errors, assumed to be independent, are then quadratically summed.

The validity of the coincidence is tested by requiring $|\Delta t^{ij}|$ to be smaller than $\Delta t^{ij}_{\text{max}} + \eta_{\text{timing}}^{\text{loose}} \times \Delta t_{\text{RMS}}^{ij}$ with $\eta_{\text{timing}}^{\text{loose}}$ being a tunable positive parameter. Finally, multi-fold coincidences require that all participating pairs of interferometers are valid. Despite the apparent weakness of condition (\ref{eq:loose}), taking into account all the delays available between detectors (a redundant set of informations) should nevertheless strongly cut false alarm events in the network considered.

\subsubsection{Tight compatibility}

If the source location in the sky is now a priori known, the compatibility condition can be tightened as the true delays $\Delta t^{ij}_{\text{true}}$ between any pair interferometers are directly computed from Eq. (\ref{eq:delay}). In this favorable case, the test requires the residual delay $| \Delta t^{ij} - \Delta t^{ij}_{\text{true}} |$ to be smaller than $\eta_{\text{timing}}^{\text{tight}} \times \Delta t_{\text{RMS}}^{ij}$, with $\eta_{\text{timing}}^{\text{tight}}$ to be tuned as well to maximize the detection efficiency at a given false alarm rate.

\subsection{Simulation procedures}

Consecutive outputs of any burst search filter are highly correlated as the input data segments strongly overlap. Thus, algorithm outputs cannot be considered as statistically independent realizations of the same random variable: one often finds clusters of consecutive data exceeding a given threshold \cite{pradier} which, in the case of a real GW burst, all correspond to the same signal. So, one has to redefine the event concept by counting only one single trigger when a consecutive set of filter output values are above the threshold. The two next paragraphs aim at giving some details on the Monte-Carlo simulation procedure for both coincidence and coherent analysis. Indeed, the latter method is more straightforward as all chunks of data are 'merged' in a precised way.

\subsubsection{Coincidences}

In the coincidence analysis simulations, an event is defined as a triplet of data: the maximum filter output, its associated time and the label of the interferometer in which it occurred. The coincidence ROC shown in this article have been constructed by using two different simulation steps: one computing the false alarm rate $\tau_{\text{norm}}$, the other estimating the detection efficiency in the various network configurations. In both cases, the compatibility between alarms in different detectors is tested according to the prescriptions given in Section \ref{subsection:coincidence_analysis} above. Results for loose and tight coincidence analyzes are presented in Section \ref{section:coincidences}.

\begin{figure}
\centerline{\epsfig{file=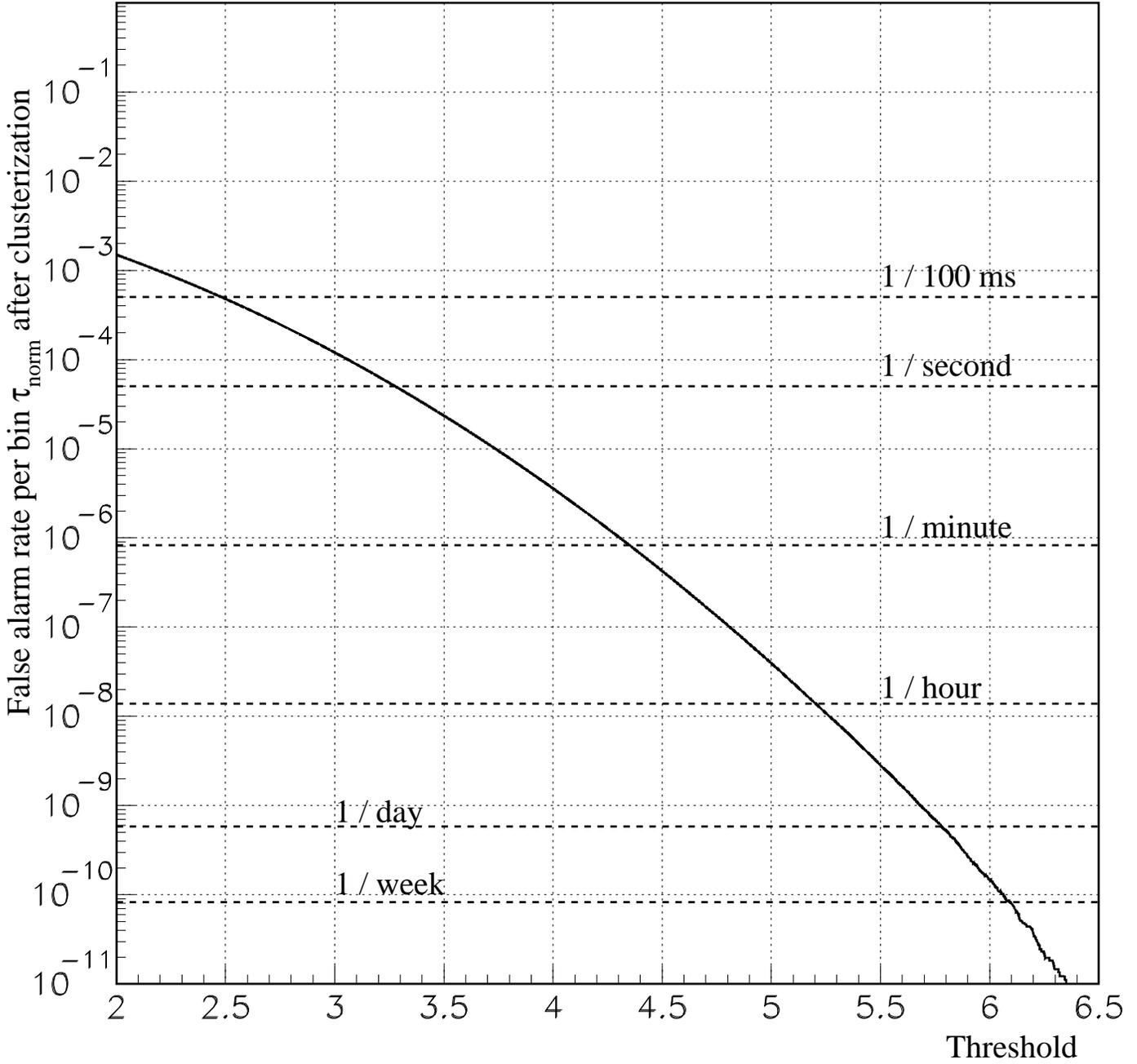,width=20cm}}
\caption{Normalized false alarm rate $\tau_{\text{norm}}$ as a function of the triggering threshold $\eta$. This curve is computed for a window size ${\mathbb{N}}=1024$~bins, i.e. 51.2~ms at a 20 kHz sampling frequency. As soon as the threshold is high enough to ensure that the probability of having a false alarm in the analysis window is below one (i.e. for a threshold around 3), this curve is completely independent on the value of $\mathbb{N}$ used in the simulation.}
\label{fig:repart_maxima}
\end{figure}

Concerning false alarms, a two-step process is used in order to limit the computing time needed for simulation. First, the rate of (clustered) false alarms as a function of the triggering threshold is computed for the single interferometer case. Figure \ref{fig:repart_maxima} shows in this case the evolution of $\tau_{\text{norm}}$ versus the threshold in a given window and some horizontal dashed lines translate $\tau_{\text{norm}}$ into more convenient values. Given the value of the threshold $\eta$, this curve is used to generate random false alarms in a particular detector with a uniform time distribution. Finally, coincidences are searched in the lists of events associated with the different detectors in the network. 

\begin{figure}
\centerline{\epsfig{file=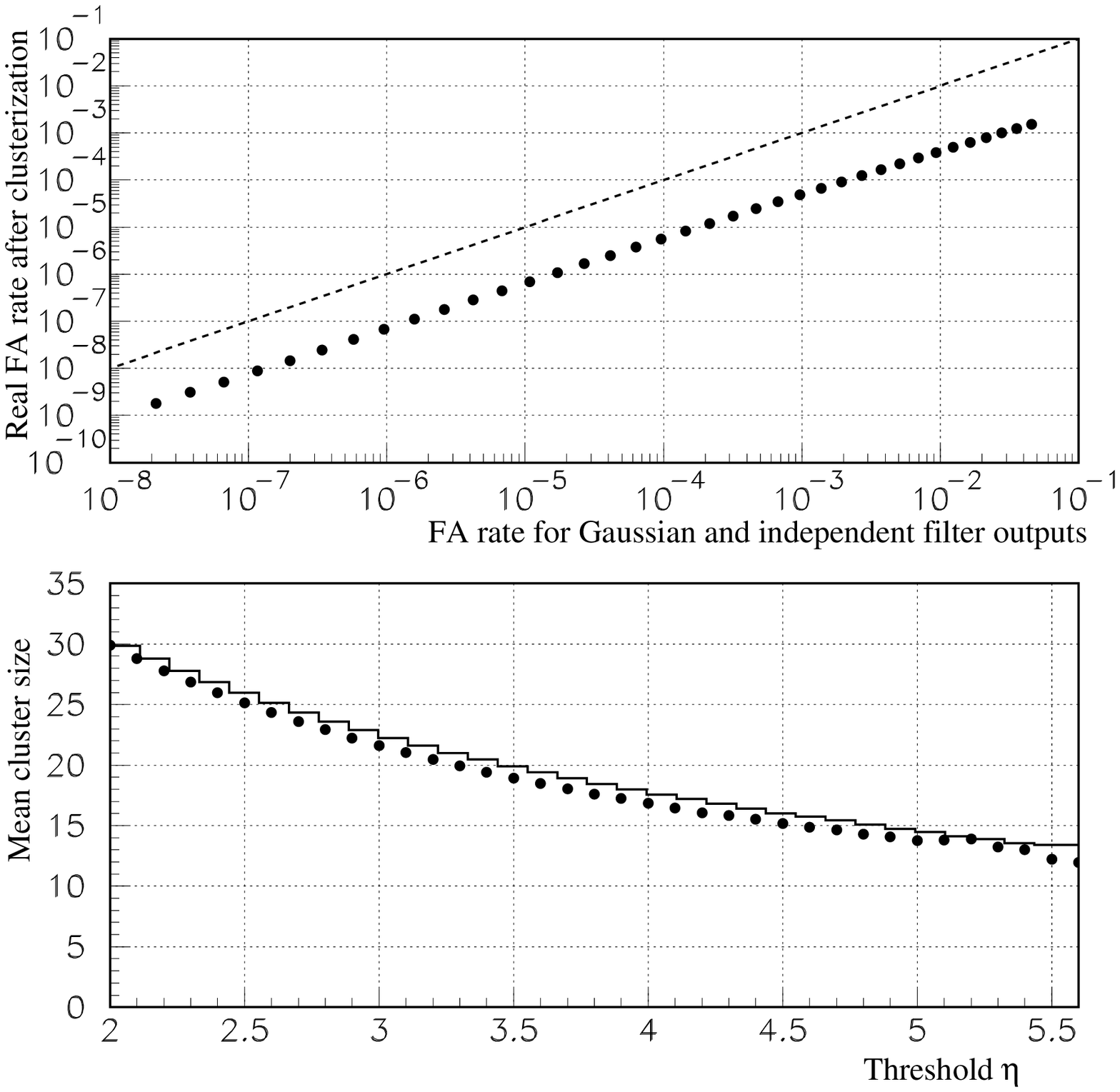,width=20cm}}
\caption{Effect of the consecutive filter output correlation on the false alarm rate. The top graph compares the normalized false alarm rate $\tau_{\text{norm}}$ for a single detector -- as computed with the simulation procedure described in the core of the paper -- with the false alarm rate $\tau_{\text{indep}}$ computed for the same threshold, assuming that successive filter outputs are uncorrelated normal random variables. The real false alarm rate is at least one order of magnitude below the Gaussian estimate, on the whole range of threshold considered. The continuous line in the bottom plot shows, as a function of the threshold $\eta$, the mean size (in bins) of the false alarm clusters which is, to a very good approximation, equal to the ratio $\tau_{\text{indep}} \, / \, \tau_{\text{norm}}$ represented by the black bullets.}
\label{fig:fa_comparison}
\end{figure}

Another point worth being mentioned is that the main effect of the alarm clustering procedure is to strongly reduce the false alarm rate $\tau_{\text{norm}}$ with respect to its estimator based on the assumption that consecutive filter outputs are independent. As shown in Figure \ref{fig:fa_comparison}, the ratio between the latter quantity and $\tau_{\text{norm}}$ is always above 10 and is indeed equal to the mean size of the false alarm clusters.

To compute the detection efficiency, the first part of the simulation changes. As the maximum delay between two existing interferometers is 41.7 ms and as millisecond bursts are considered, synchronized data windows of ${\mathbb{N}}=1024$~bins (corresponding to 51.2 ms for the sampling frequency $f_{\text{samp}}=20$~kHz) are enough to contain all the signal components after interaction with the detectors. Then, events are searched for in the data chunks from different detectors and the tests of coincidence compatibility between alarms are performed as in the false alarm case.

As false alarm rates are kept low, the noise realizations are in this case required not to produce any false alarm. Of course, this bias is important only for large values of $\tau_{\text{norm}}$, let say above $10^{-5}$. Below, the probability to have a false alarm in ${\mathbb{N}}$ data is under 1\%, and so the noise bias does not play any significant r\^ole. In this way, one is sure that all triggers are due to some signal components and that coincidence efficiencies are not affected by false alarm contributions. 

\subsubsection{Coherent analysis}

The coherent analysis simulation is much simpler as the interferometer filter outputs are merged in a single data flow. Assuming a fixed analysis window size (${\mathbb{N}}=1024$ for instance), false alarm rates and detection efficiencies are simply equal to the ratio of the number of simulations with outputs exceeding the threshold to the total number of simulations. The only assumption made is that errors in the source location are negligible, so that the relative shifts applied to synchronize the interferometer data segments are exact.

\section{Single interferometer study}
\label{section:single}

This section summarizes the performances of a single interferometer, both in term of detection efficiency and of timing resolution, providing the necessary inputs for the coincidence studies.

\subsection{ROC}

\begin{figure}
\centerline{\epsfig{file=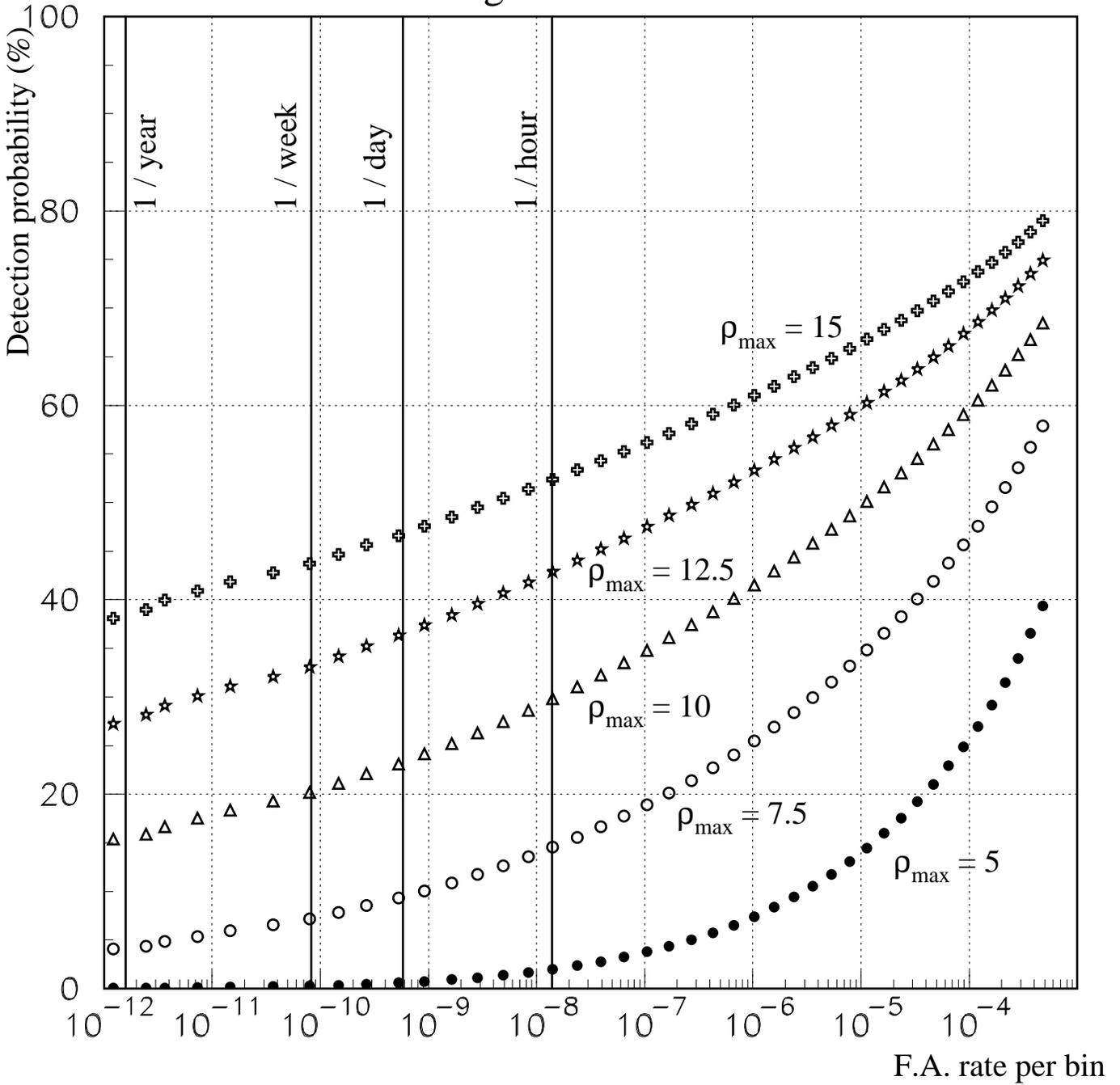,width=20cm}}
\caption{ROC for a single interferometer with $\omega = 1$ ms. Five curves corresponding to different values of the maximal SNR -- $\rho_{\text{max}}=5$, 7.5, 10, 12.5 and 15 -- are plotted. Like for all the forthcoming ROC, vertical lines give convenient conversions of the false alarm rate per bin, assuming a sampling frequency $f_{\text{samp}}=20$ kHz: from left to right, one false alarm per year, per week, per day, and per hour respectively. Comparing these curves clearly show that one cannot have a high detection efficiency associated to an high confidence level -- i.e. a very small false alarm rate -- by using only one detector to seek GW bursts.}
\label{fig:roc_single}
\end{figure}

Figure \ref{fig:roc_single} collects some ROC for a single interferometer -- the detection efficiency $\epsilon$ versus the normalized false alarm rate $\tau_{\text{norm}}$. Five different values of the optimal SNR $\rho_{\text{max}}$ are considered, ranging from 5 to 15. Like for all the similar plots of the article, some particular values of the normalized false alarm rate converted in more practical units are represented by vertical lines: from 1/year to 1/hour. This false alarm range should cover all interferometer operating configurations. 

Assuming $\rho_{\text{max}}=10$ and an interferometer optimally orientated would lead to a detection efficiency very close to 100\% in the whole false alarm range. Unfortunately, because of the non-uniform antenna pattern, the detected signal amplitude is strongly reduced on average, and so the probability of detection. Thus, for intermediate values of $\rho_{\text{max}}$, the detection is not likely in a single detector, provided that the trigger threshold is kept high enough to have a small false alarm rate. Let us consider for instance the curve corresponding to $\rho_{\text{max}}=10$, a typical value one can expect for a supernova at the Galactic center \cite{ZM,DFM}. In order to reach a 50\% efficiency, the detector must be run at a false alarm rate of roughly 1/second!

\subsection{Timing performances}
\label{subsection:timing_performances}

As all compatibility tests for network data analysis methods are based on time delays between the different interferometer candidates, the timing resolution of a filter is another important quantity. Like for the detection problem, the matched filter appears to have the best resolution, as shown in \cite{hello_timing} where optimal and sub-optimal filtering methods are compared. A first study of the Wiener filtering for a Gaussian signal in \cite{arnaud_burst_2} showed that the timing resolution (i.e. the RMS of the difference $\Delta t$ between the timing of the maximum filter output and the real GW timing) could be simply parameterized:

\begin{equation}
\Delta t_{\text{RMS}} \; \approx \; 0.15 \text{ ms } \, \left( \, \frac{ \omega }{ 1 \, \text{ms} } \, \right) \, \left( \, \frac{ 10 }{ \rho } \, \right)
\label{eq:timing_resolution}
\end{equation}
with an excellent agreement between the fit and the real RMS as soon as $\rho \ge 6$.

Here, we extend this work by studying the evolution of $\Delta t_{\text{RMS}}$ on a larger range of $\rho$: from 0 (no signal) to a very large value -- see Figure \ref{fig:timing_resolution}. This result will be used later to validate coincidences in Section \ref{section:coincidences}.

As we focus in this article on a Gaussian peak with width $\omega=1$~ms lasting around 6 ms in total, a window of ${\mathbb{N}}=512$ has been chosen to compute the evolution of $\Delta t_{\text{RMS}}$ versus $\rho$. Indeed, it corresponds to 25.6~ms for a 20 kHz sampling frequency, a duration large enough to include the whole signal. Choosing a much larger value for $\mathbb{N}$ is not suitable as for a negligible GW signal completely dominated by the noise, $\Delta t$ is uniformly distributed in the analysis window. The timing error RMS would then grow 'artificially' with $\mathbb{N}$. Consequently, the compatibility condition would be more easily satisfied, leading thus to an increase of the false alarm rate. On the other hand, for very large values of $\rho$, one expect $\Delta t_{\text{RMS}}$ to scale like $1/\sqrt{\rho}$, i.e. a slower variation than Eq. (\ref{eq:timing_resolution}). Ultimately, the timing resolution would be limited by the sampling frequency.

\section{Coincidences}
\label{section:coincidences}

In this section, different configurations of interferometric detector networks are studied: the 3-interferometer network Virgo-LIGO, the LIGO 3-interferometer network including the Hanford 2-km detector, and the full network made of the six first generation interferometers. They allow one to predict the performances that could be achieved in detecting GW bursts in coincidence in the future. This network detection model could easily be updated when final relative sensitivities are known, or when new second generation detectors appear. In these studies, the loose compatibility test is always used, apart in the last paragraph \ref{subsection:tight_coincidences} where a summary of tight coincidence performances is presented.

\subsection{Virgo-LIGO network}

First, we consider the three-interferometer network Virgo-LIGO. Studying its efficiency is important for two main reasons: it includes the detectors with the best foreseen sensitivities and a threefold detection is the minimum number of coincidences required to estimate the source location in the sky.

\subsubsection{Two-interferometer coincidences}

The simplest network is made of two interferometers. Its performances depend a lot of the particular configuration considered, as shown in the following. For the timing compatibility condition, three different values of $\eta_{\text{timing}}^{\text{loose}}$ have been tested in simulations: 1, 2 and 3. It turns out that the best compromise between low false alarm rate and high detection efficiency is obtained with $\eta_{\text{timing}}^{\text{loose}}=1$, value used in the following for all loose coincidence tests. 

\begin{figure}
\centerline{\epsfig{file=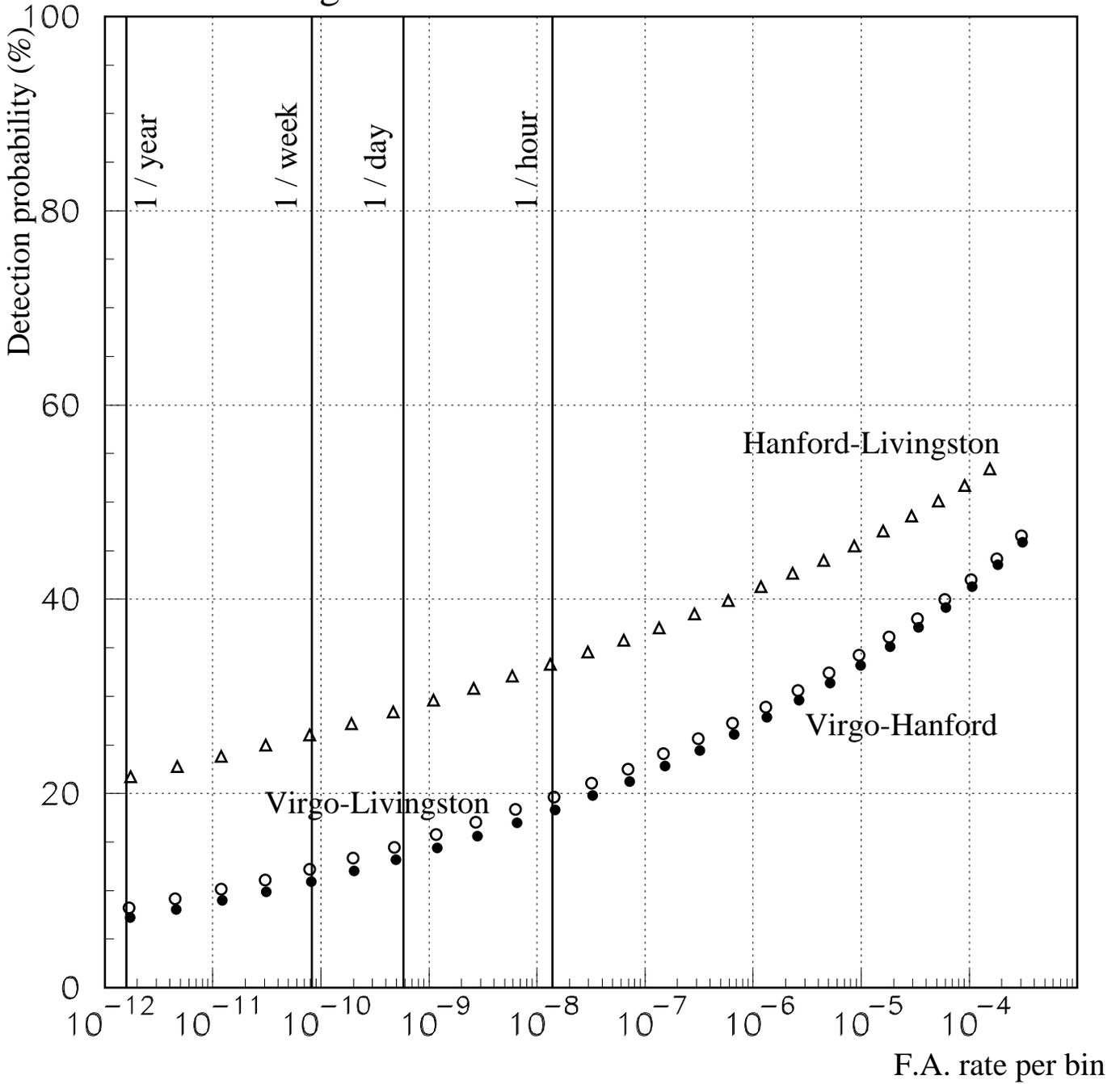,width=20cm}}
\caption{ROC comparing the pairs of detectors belonging to the three-interferometer network Virgo + the two 4 km-LIGO detectors. In this graph, the maximum SNR $\rho_{\text{max}}$ is set to 10. As the two LIGO interferometers have been built together in order to be 'aligned', this network shows better performance than the two other ones.}
\label{fig:roc_vhl_2detectors_comparison}
\end{figure}

Figure \ref{fig:roc_vhl_2detectors_comparison} compares the ROC computed for the three pairs of detectors with $\rho_{\text{max}}=10$. The first point to notice is that the efficiency never reaches 60\%, even at very high false alarm rates: two interferometers are not enough to guarantee a likely detection of such bursts. Then, the configuration associating the two LIGO detectors shows clearly better performances than the two others made of Virgo plus one LIGO. Two reasons explain these differences:

\begin{itemize}
\item The two LIGO interferometers have been built in order to maximize the correlation between their antenna patterns, increasing the coincidence efficiency. This is the dominant effect -- see the next paragraph.
\item The LIGO detectors are close with respect to Virgo -- see Table I in Section \ref{subsection:coincidence_analysis} -- and less random coincidences are allowed in the compatibility window. So, the false alarm rate is shifted to the left thanks to this effect.
\end{itemize}

This feature remains true if one compares all pairs of interferometers chosen among the full network of six detectors: the two LIGO configuration ROC is clearly better than any other. Even the Virgo-GEO600 pair -- the two closest instruments -- cannot compete: for a given threshold, the false alarm rate is lower, but also the efficiency as the angular patterns do not overlap well. 

\begin{figure}
\centerline{\epsfig{file=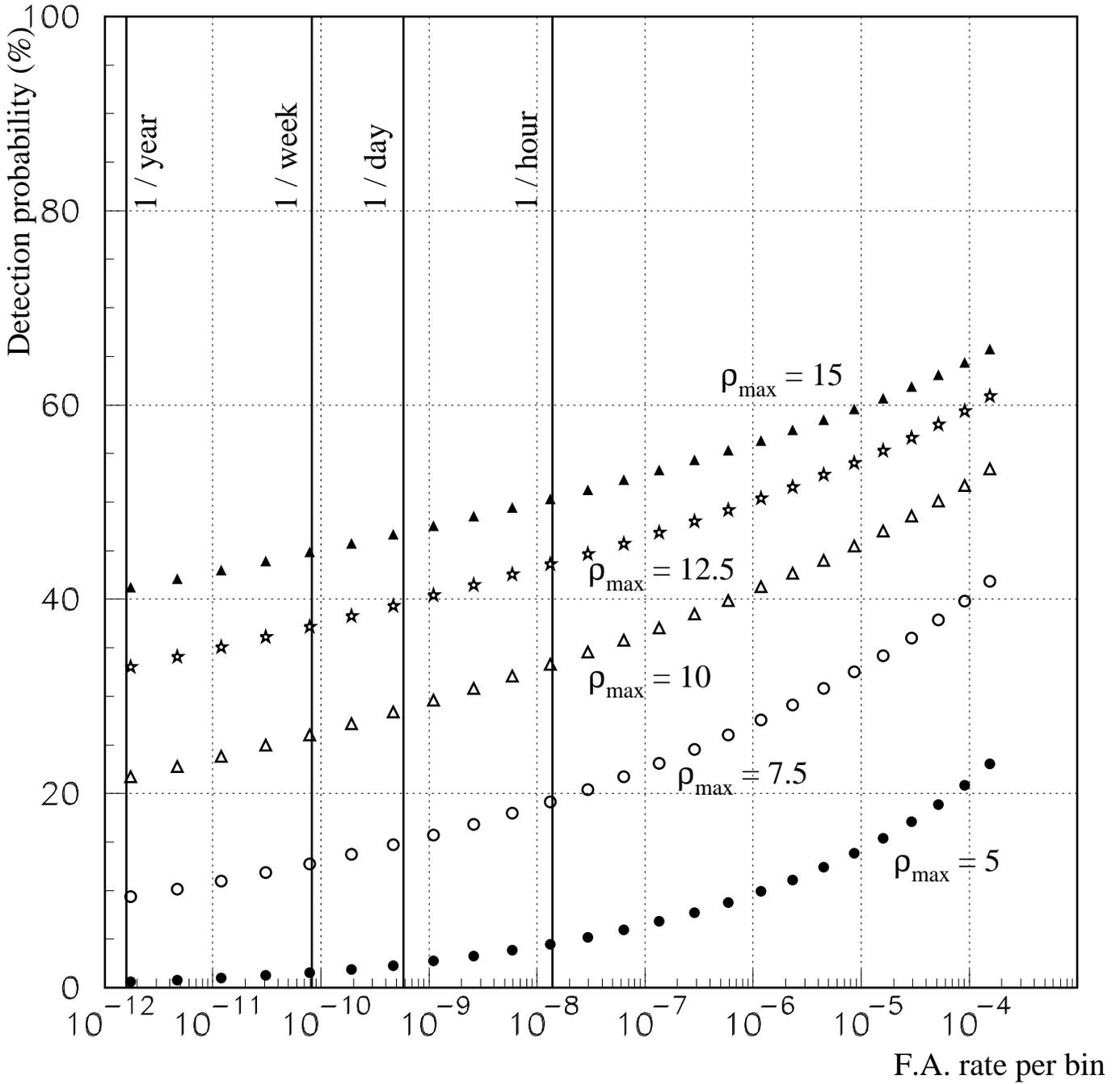,width=20cm}}
\caption{ROC characterizing the two LIGO detectors which form the best pair of interferometers. The five curves cover the range between $\rho_{\text{max}}=5$ and 15. Even for the larger values of the optimal SNR, detection efficiency remains below 50\% for manageable false alarm rates (below 1/s).}
\label{fig:roc_LIGO}
\end{figure}

Figure \ref{fig:roc_LIGO} shows how the two LIGO detection efficiency evolves for different values of the optimal SNR, between 5 and 15. The larger $\rho_{\text{max}}$, the better the efficiency, but the improvement remains limited: even for $\rho_{\text{max}}=15$, the detection probability is only around 50\% for $\tau_{\text{norm}}=1$/ hour. So, a two detector network appears to be not sufficient. Yet, a last point to be mentioned is that the efficiency decreases more slowly with the false alarm rate than for the single detector case -- see Figure \ref{fig:roc_single} for comparison. The next section will clearly show the interest of this behaviour. 

\subsubsection{Coincidence strategy comparison}

\begin{figure}
\centerline{\epsfig{file=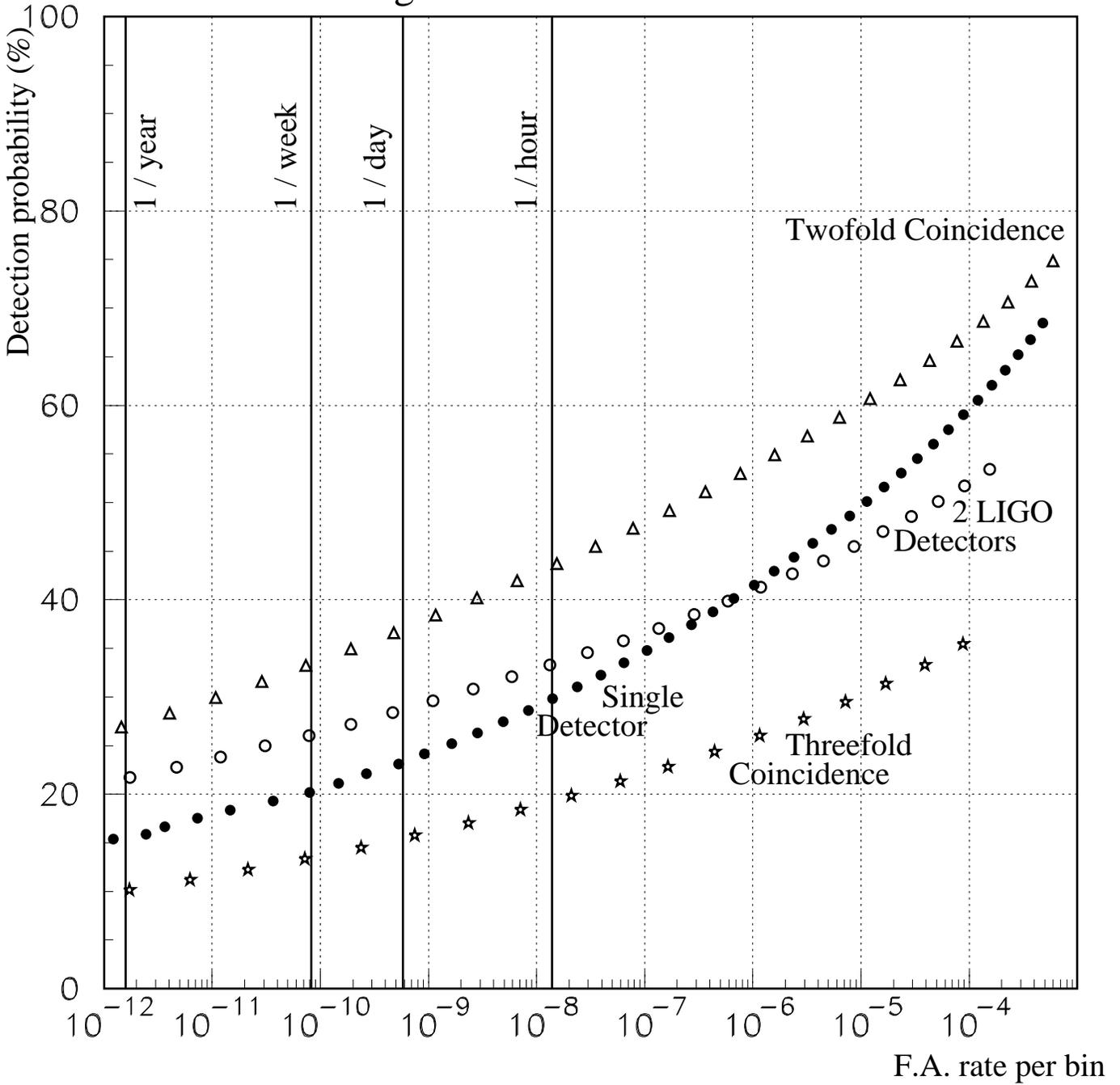,width=20cm}}
\caption{ROC comparing all the coincidence strategies in the Virgo-LIGO network for $\rho_{\text{max}}=10$. The best configuration requires at least two detections among three, but its efficiency remains limited for a false alarm rate small enough; in addition, the full coincidence appears unlikely. These two results clearly show that larger networks are required. A last point worth being mentioned is that for manageable false alarm rates, the detection efficiency is better for a two interferometer network than for a single detector.}
\label{fig:roc_coincid_vhl}
\end{figure}

Figure \ref{fig:roc_coincid_vhl} compares for $\rho_{\text{max}}=10$ all the possible coincidence strategies in the Virgo-LIGO network: single detector, coincidences in the two LIGO interferometers (the best pair of detectors), twofold coincidences (at least two detections among three) and finally full coincidences. The twofold coincidence strategy is clearly the best: its ROC is above the other ones in the full range of false alarms covered by the graph. Yet, it does not show very high efficiencies: only a bit more than 40\% for 1 false alarm per hour, and around 25\% at the level of 1 per year. In addition, threefold coincidences are very inefficient, and these results are indeed similar for all triplets of interferometers. The networks with the best ROC all include the two LIGO detectors, but their efficiencies depend weakly on the location of the third one on Earth: replacing Virgo by GEO600 gives a slightly better result, while using TAMA300 or ACIGA decreases a bit the efficiency at given false alarm rates. Therefore, a three interferometer network does not appear large enough to reach high detection efficiencies.

Two other interesting points can be extracted from the plots. First, as soon as the false alarm rate becomes small enough to be realistic (say around $\tau_{\text{norm}}=1/$~few seconds), the ROC corresponding to coincidences between the two LIGO detectors crosses the single interferometer ROC and shows larger efficiencies at fixed false alarm rate. This crossing is due to the fact that the large decrease in the false alarm rate due to the compatibility condition required for a LIGO coincidence is stronger than the corresponding loss in efficiency caused by the twofold detection. This demonstrates that searching GW bursts in one detector is not only discarded by the small confidence level one can associate to such events, but also because this method is simply less efficient than others.

Second, for smaller false alarm rates, the ROC for twofold coincidences and for the two LIGO get closer. This is due to the fact that the two LIGO antenna patterns are close one each other and quite different from the Virgo pattern. Therefore, when the threshold increases, coincidences between Virgo and one LIGO are more strongly suppressed than for the LIGO pair. Yet, the two curves are widely separated in the whole range of false alarm: adding Virgo allows one to improve significantly the efficiency with respect to the two LIGO detectors alone -- more than 30\% on a relative scale. So, going from two to three interferometers in the network is a clear improvement.

\subsection{Full network of six interferometers}

\begin{figure}
\centerline{\epsfig{file=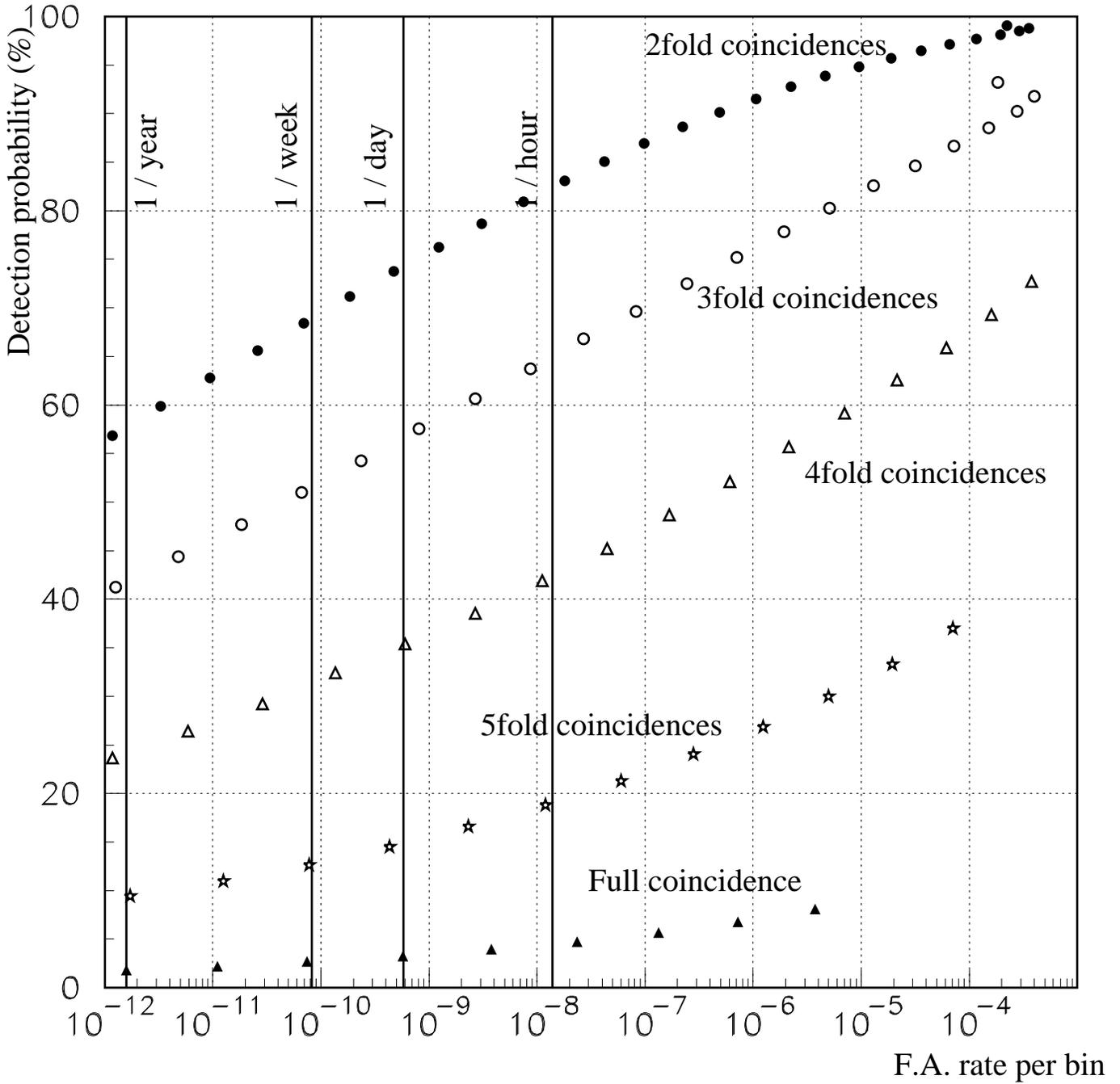,width=20cm}}
\caption{ROC for the full network coincidence strategies, from twofold (top curve) to sixfold coincidences (bottom curve) and with $\rho_{\text{max}}=10$. Detection efficiencies clearly improve by going to three to six interferometers; indeed, in this larger configuration, both twofold and threefold coincidences are likely, even at very low false alarm rates.}
\label{fig:roc_coincid_full}
\end{figure}

As the previous section pointed out that a three interferometer network is not promising enough, we compute in this section the detection efficiency for the full network of six detectors. Operating such configuration in a near future should be a clear goal of the worldwide GW community. Indeed, this network reaches quite promising detection efficiencies, as Figure \ref{fig:roc_coincid_full} shows. The ROC, computed again for $\rho_{\text{max}}=10$, correspond to coincidence strategies in which a minimal number of detections is required: from two (top curve) to six (bottom curve). In this configuration, twofold coincidences are quite likely: more than 80\% for 1 false alarm per hour and still about 60\% for 1/year. In addition, threefold coincidences appear possible: in this case, the efficiency is around 60\% for 1 false alarm per day. On the other hand, higher multiplicity coincidences are less and less efficient.


Table~II summarizes the loose coincidence results previously presented. It collects detection efficiencies from various strategies, sampled at representative false alarm rates.

\begin{center}
\begin{tabular}{|c|c|c|c|c|c|}
\hline Configuration & $\tau_{\text{norm}}=1$/year & $\tau_{\text{norm}}=1$/week & $\tau_{\text{norm}}=1$/day & $\tau_{\text{norm}}=1$/hour  \\
\hline Single detector & 16\% & 20\% & 23\% & 30\% \\
\hline At least 2/3 (Virgo-LIGO network) & 27\% & 34\% & 37\% & 44\% \\
\hline 3/3 (Virgo-LIGO network) & 10\% & 13\% & 15\% & 20\% \\
\hline At least 2/6 & 57\% & 69\% & 74\% & 83\% \\
\hline At least 3/6 & 47\% & 52\% & 57\% & 65\% \\
\hline At least 4/6 & 23\% & 31\% & 35\% & 42\% \\
\hline
\end{tabular}
\end{center}
\vskip -0.2truecm
\centerline{Table~II: Loose coincidence efficiency comparison for $\rho_{\text{max}}=10$}

\subsection{The LIGO network}
\label{subsection:LIGO_network}

The LIGO system actually consists of three interferometers: the two 4-km detectors in Hanford and Livingston, and the Hanford 2-km interferometer located in the same vacuum tube as the larger instrument. Assuming the most optimistic situation in which the dominant noises of these two neighbor detectors are independent, their close locations significantly reduce the number of random false alarms between them. On the other hand, the difference in the arm lengths reduces the 2-km detector sensitivity by a factor two. Therefore, it is interesting to see how these two effects balance and ROC are well-suited for such a study.

\begin{figure}
\centerline{\epsfig{file=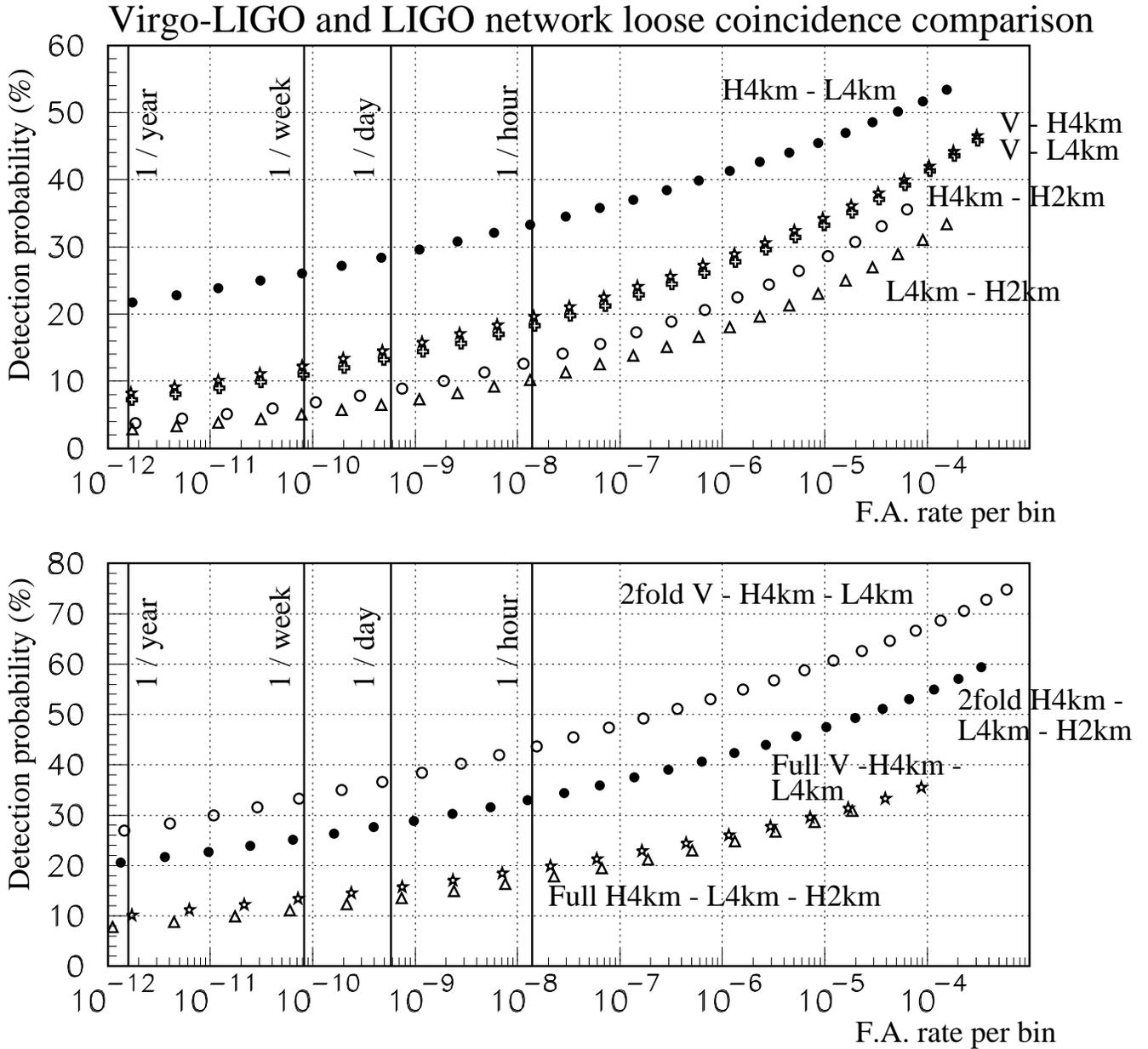,width=18cm}}
\caption{Comparison between two networks of three interferometers: Virgo and the two LIGO 4-km detectors on the one hand, and the three LIGO detectors (including the LIGO-Hanford 2-km) on the other hand. In order to simplify the labels of the two plots, the interferometer names are shortened: Virgo (V), LIGO Hanford 4-km (H4km) and 2-km (H2km) and LIGO Livingston 4-km (L4km). The top graph presents ROC corresponding to coincidences between pairs of detectors: from top to bottom, the two LIGO 4-km detectors, Virgo associated with each of the two LIGO 4-km interferometers, the two Hanford detectors (4 km and 2 km) and finally Livingston 4-km with Hanford 2-km. The bottom graph compares the two possible strategies involving all the detectors of these networks: twofold coincidences (at least two detections among three) and full coincidences. As the two LIGO Hanford detectors have identical locations, their coincidence false alarm rate is lower than for any other pair of detectors. Yet, this does not compensate the difference in sensitivity between them which limits their detection efficiency: the full LIGO network is less efficient than the Virgo-LIGO detector. This clearly shows the importance of the final interferometer sensitivity.}
\label{fig:roc_LIGO_ntwrk_comparison}
\end{figure}

Figure \ref{fig:roc_LIGO_ntwrk_comparison} compares the full LIGO network with the Virgo-LIGO (4-km) network. The top plot presents ROC for pairs of interferometers: the two LIGO 4-km detectors, Virgo and each of the LIGO 4-km interferometers, and finally each of the LIGO 4-km interferometers with the Hanford 2-km detector. The bottom plot compares the coincidence strategies involving the three detectors in each network: twofold coincidence (at least two detections among three) and the full coincidence.

These graphs show that the performance of the full LIGO network is worse than the Virgo-LIGO (4 km) network: detections efficiencies at given false alarm rate are better in the latter case. The reduction factor of the LIGO Hanford 2-km detector sensitivity plays a more important r\^ole than the gain in false alarm rate provided by the coincident locations of the Hanford interferometers. Conversely, this is a strong indication that adding in a network detectors less sensitive than others will only give limited improvements in detection efficiency.

\subsection{Tight coincidences}
\label{subsection:tight_coincidences}

\begin{figure}
\centerline{\epsfig{file=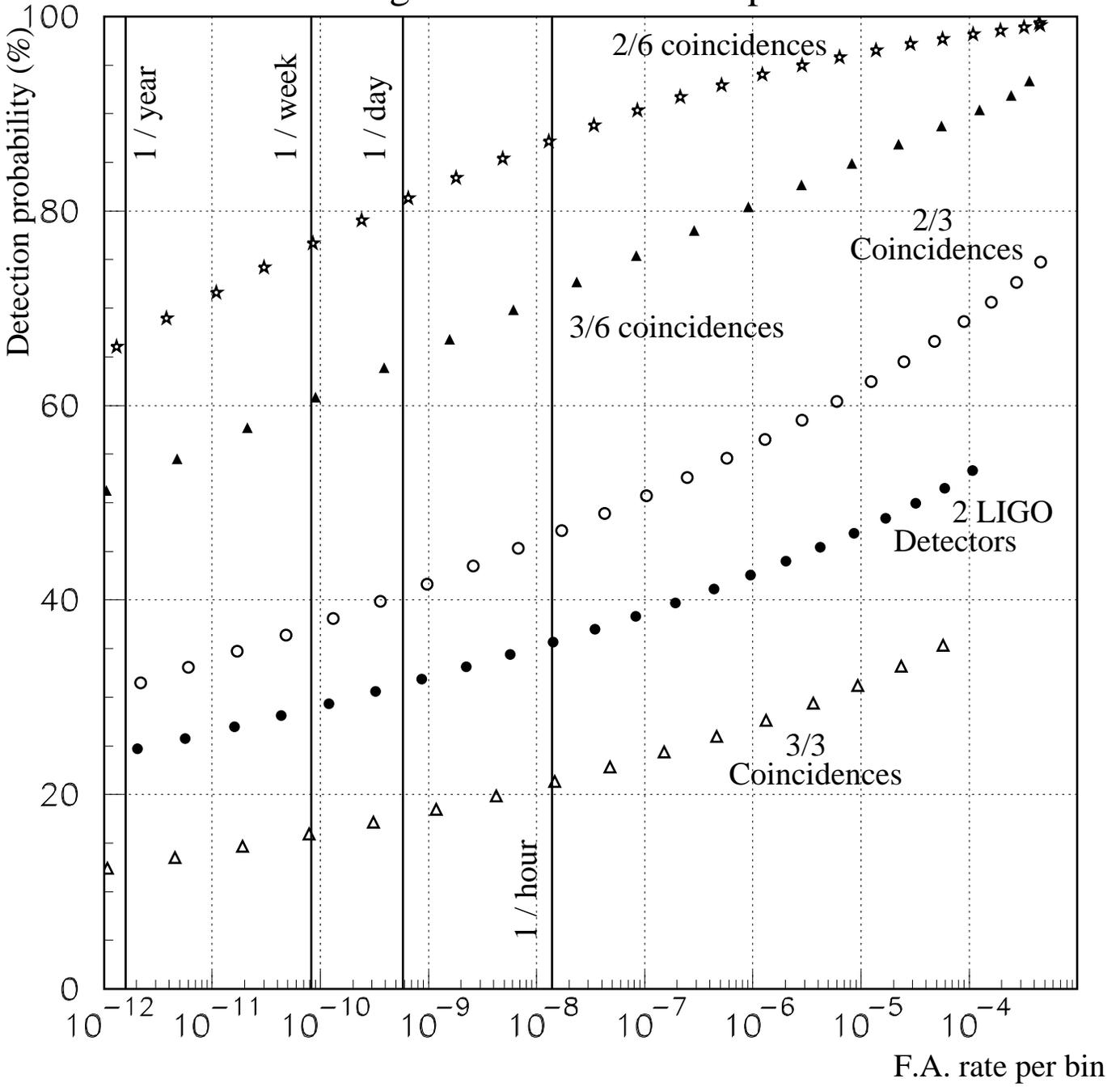,width=20cm}}
\caption{Comparison of ROC ($\rho_{\text{max}}=10$) corresponding to various tight coincidence strategies: LIGO coincidences, twofold and threefold detections in the Virgo-LIGO network and in the full network of six interferometers.}
\label{fig:tight_comparison}
\end{figure}

To conclude this section dealing with coincidence detection, Figure \ref{fig:tight_comparison} compares ROC computed for different tight coincidence strategies: coincidences between the two 4-km LIGO detectors, twofold and threefold coincidences in both the Virgo-LIGO network and the full network of six interferometers. As the source location is assumed to be known here, such curves can be directly compared with the coherent analysis results presented in the next section.

As for the loose coincidence case, a tuning of the parameter $\eta_{\text{timing}}^{\text{tight}}$ has been performed; values ranging between 1 and 5 have been used, with the choice $\eta_{\text{timing}}^{\text{tight}}=3$ giving the best ROC. Comparing Figure \ref{fig:tight_comparison} with Figures \ref{fig:roc_vhl_2detectors_comparison} to \ref{fig:roc_coincid_full} -- corresponding to the loose compatibility criterion -- shows that the tight coincidence ROC present larger relative efficiencies. Yet, the improvement remains limited, from 10 to 20\% in relative. Therefore, the main limitation of the coincidence ROC appears to be the coincidence algorithm itself, handling only binary informations (signal present or absent) in the different detectors with a fixed threshold.

\section{Coherent analysis}
\label{section:coherent_analysis}

Compared to coincidence searches, coherent analysis methods use a more complete set of informations coming from the different components of the network; thus, their detection efficiency is larger. On the other hand, using them in real analysis requires an additional hypothesis on the source location, which allows one to properly shift the various detector outputs to synchronize them. In the most general case, the source position in the sky, monitored by two angles, is unknown and thus must be added to the set of unknown parameters describing the GW signal. So, many templates must be run in parallel to ensure an efficient coverage of the sky, each of them focusing on a particular area. Consequently, any comparison between coincidence and coherent methods must take into account this fact, leading in particular to a renormalization of the false alarm rate in the case of loose coincidences -- see Section \ref{section:comparison} for more details. Yet, if the source location is already known (e.g. from informations given by detectors sensitive to other radiations), this restriction is lifted and one coherent algorithm is enough. In this case, its performances can be directly compared with the tight coincidence scenario.

To cope with this requirement, the study of coherent analysis methods is performed in two steps. In this section, ROC are computed for a single coherent algorithm, assuming a perfect knowledge of the source location in the sky. Then, the number of filters needed to cover the sky is estimated in Section \ref{section:sky_tiling}, using the formalism of Ref. \cite{Owen}.

\subsection{Derivation of the statistics from the likelihood ratio}
\label{analysis:deriv_stat}

In the following, the derivation of the coherent statistics based on the likelihood ratio is briefly recalled. Assuming known the source sky location, the template to be used in the $i$-th detector $D^i$ takes the form

\begin{equation}
s^i(t) \; = \; \underbrace{F_i}_{\text{beam pattern term}} \; \times \; \underbrace{K}_{\propto \, 1 / \text{distance}} \; \times \; s^0\left( t \, - \, \delta t^i \right)
\end{equation}
where $s^0$ is the generic template shifted by the time delay $\delta t^i$ and scaled by $F_i$, the factor giving the quality of the interaction between the antenna and the GW. Without loss of generality, one can assume that $\langle s^0 \, | \, s^0 \rangle (0) = 1$.

Let us consider first the search of a known signal in a single detector output. The most efficient method is in this case the Wiener filter. Its expression naturally arises in the framework of the likelihood ratio, defined as the conditional probability to have a particular set of data assuming that the signal is present -- see e.g. the corresponding discussion in \cite{Finn}.

By using the same method, one can define a global likelihood ratio for a set of detectors. With the hypothesis made on the interferometer noises (Gaussianity and independence), the logarithm of the 'network' likelihood ratio $\ln\lambda$ is computed by simply summing the corresponding contributions for the $P$ single detectors $D^i$:

\begin{equation}
\ln\lambda \; = \; \sum_{i=1}^{P} \, \ln\lambda^i
\nonumber
\end{equation}

Its final expression can be found in Ref. \cite{Finn,PDB}:

\begin{equation}
\ln\lambda \; = \; \sum_{i=1}^{P} \; \frac{ \langle s^i \, | \, x^i \rangle }{ \sigma_i^2} \; - \; \frac{ 1 }{ 2 } \; \sum_{i=1}^{P} \; \frac{ \langle s^i \, | \, s^i \rangle }{ \sigma_i^2} \;
\end{equation}
which is a second order polynomial function in the unknown $K$. Still following \cite{PDB}, maximizing over this variable and extracting the polarization angle $\psi$ from the beam pattern functions gives the following expression:

\begin{equation}
\ln\lambda \; = \; \frac{1}{2} \; \left[ \frac{ \left( \cos2\psi \, \vec{A} \, + \, \sin2\psi \, \vec{B} \right) \, . \, \vec{\gamma} }{ || \cos2\psi \, \vec{A} \, + \, \sin2\psi \, \vec{B} || } \right]^2 
\end{equation}
with: 

\begin{equation}
\nonumber
A^i \; = \; \frac{ a^i }{ \sigma^i } \;\;\;\; B^i \; = \; \frac{ b^i }{ \sigma^i } \;\;\;\; \text{and} \;\;\;\; \gamma^i \, (t_0) \; = \; \frac{ \langle s | x^i \rangle }{ \sigma^i } \;\;\;\; \text{for} \;\;\;\; i=1,...,P
\nonumber
\end{equation}

The vector $\vec{\gamma}$ can be expanded in the following way:

\begin{equation}
\vec{\gamma} \; = \; \Gamma_A \, \vec{A} \; + \; \Gamma_B \, \vec{B} \; + \; \vec{\gamma}_\perp \;\;\; \text{with} \;\;\; \vec{\gamma}_\perp \, \perp \, \vec{A} \;\;\; \text{and} \;\;\; \vec{\gamma}_\perp \, \perp \, \vec{B}
\label{eq:expansion}
\end{equation}

Therefore, only the two first terms of this sum contribute to the likelihood ratio. One can now compute the value of $\psi$ which maximizes it. By applying the Schwarz inequality, one gets the following upper bound

\begin{equation}
(\ln\lambda)^{\; \text{max}} \; \leq \; \frac{ 1 }{ 2 } \; \left\| \; \Gamma_A \, \vec{A} \; + \; \Gamma_B \, \vec{B} \; \right\|^2
\end{equation}
which is reached by choosing $2 \psi$ such as 

\begin{equation}
\cos( 2 \psi ) \; = \; \frac{ \Gamma_A }{ \sqrt{ \Gamma_A^{\;2} \, + \, \Gamma_B^{\;2} } } 
\;\;\;\; \text{and} \;\;\;\;
\sin( 2 \psi ) \; = \; \frac{ \Gamma_B }{ \sqrt{ \Gamma_A^{\;2} \, + \, \Gamma_B^{\;2} } } 
\nonumber
\end{equation}

The maximum of the likelihood ratio statistics is thus proportional to the square of the norm of the orthogonal projection of $\vec{\gamma}$ on the subspace generated by the couple $\left(\vec{A},\vec{B}\right)$. Computing the values of $\Gamma_A$ and $\Gamma_B$ allows one to give a compact expression of this new statistics,denoted $\Lambda$ in the following:

\begin{equation}
\Lambda \; = \; \frac{ \left\| \; \left( \vec{\gamma} \, . \, \vec{B} \right) \, \vec{A} \; - \; \left( \vec{\gamma} \, . \, \vec{A} \right) \, \vec{B} \; \right\|^2 }{ A^2 \, B^2 \; - \; \left( \vec{A} \, . \, \vec{B} \right)^2 }
\label{eq:Lambda}
\end{equation}
with $A = \left\|\vec{A}\right\|$ and $B = \left\|\vec{B}\right\|$.

\begin{figure}
\centerline{\epsfig{file=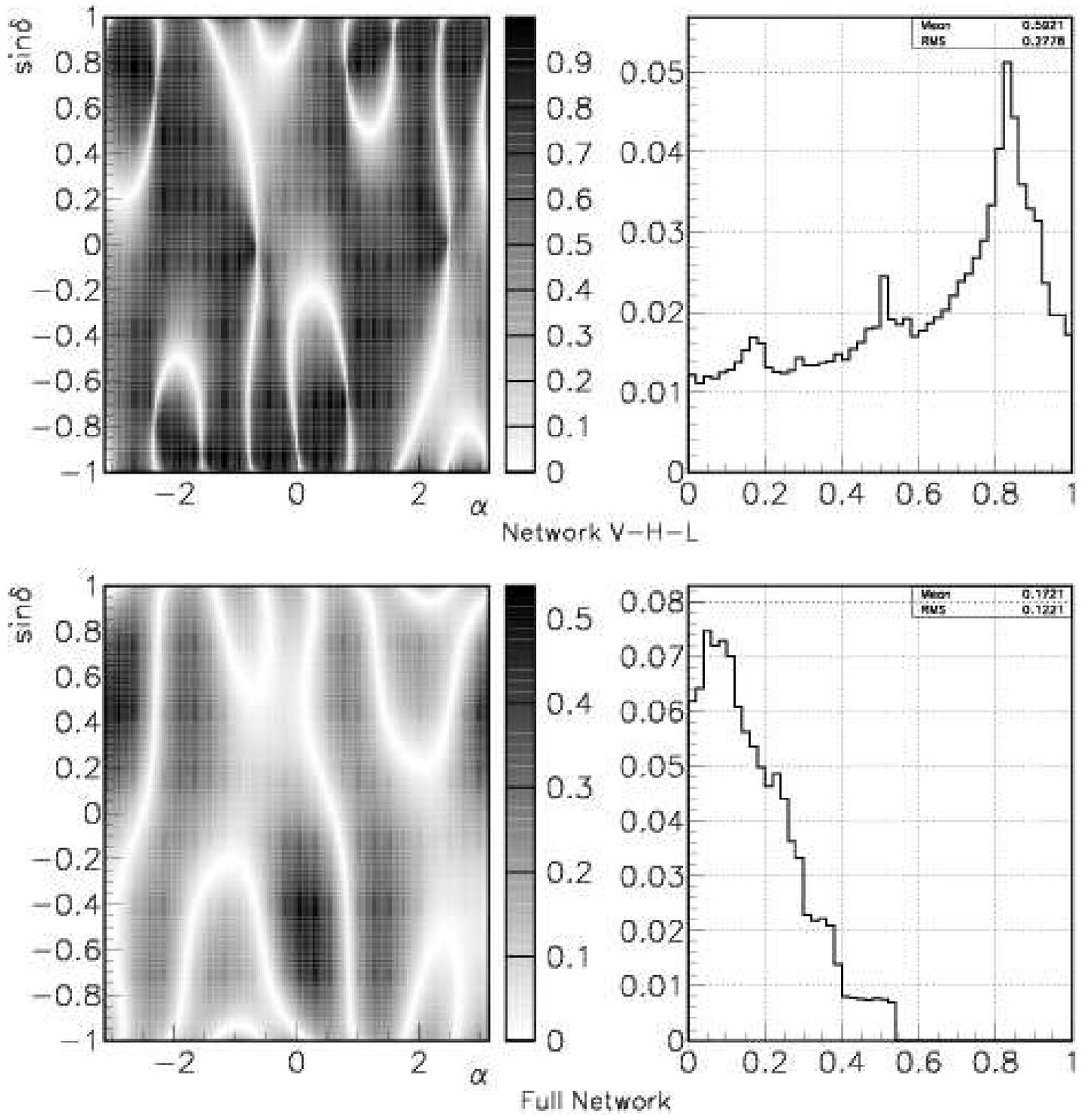,width=20cm}}
\caption{Sky maps and distributions of $|\cos\theta_{AB}|$ as a function of the source sky coordinates $(\alpha,\delta)$ for two different network configurations: (top) Virgo and the two LIGO interferometers ; (bottom) the full set of six antennas.}
\label{fig:cos_alpha_beta}
\end{figure}

This last expression shows that all the former calculations are meaningless if $\vec{A}$ and $\vec{B}$ are parallel -- indeed, invalidating Eq. (\ref{eq:expansion}). Fortunately, this critical situation is unlikely in the sky, as shown in Figure \ref{fig:cos_alpha_beta} where the sky map of $| \cos\theta_{AB} | = | \vec{A} . \vec{B} | \, / \, (A B)$ is represented as a function of the celestial sphere coordinates for the Virgo-LIGO network and for the full set of six detectors.

For the first network, the values $\cos\theta_{AB} = \pm 1$ can be reached only in a very small area of the sky while in the second case, the absolute value of the cosine remains below 0.6 in any direction of the sky. This difference is simply due to the fact that the vectors $\vec{A}$ and $\vec{B}$ have grown from 3 to 6 components and are therefore less likely to be collinear. 

\subsection{Statistical behavior of $\Lambda$}

To use the estimator $\Lambda$ for data analysis purpose, one can first study its statistical properties under the hypothesis of noise only. To do this, one has to rewrite Eq. (\ref{eq:Lambda}) on a different way. First, introducing an orthogonal basis in the plane containing $\vec{A}$ and $\vec{B}$ by defining the two following vectors \cite{PDB}:

\begin{equation}
\nonumber
\vec{u} =  \frac{ \vec{A} }{ A } \; + \; \frac{ \vec{B} }{ B } \;\;\;\; \text{and} \;\;\;\; \vec{v} =  \frac{ \vec{A} }{ A } \; - \; \frac{ \vec{B} }{ B }
\end{equation}

$\Lambda$ can then be rewritten:

\begin{equation}
\Lambda \; = \underbrace{ \frac{ A^2 \, B^2 \, || \, \vec{u} \, ||^2 \; || \, \vec{v} \, ||^2 }{ 4 \; \left[ A^2 \, B^2 \; - \; \left( \vec{A} \, . \, \vec{B} \right)^2 \right] } }_{\text{Coefficient only depending on the source sky location}} \times \; \left[ \; \left( \vec{\gamma} \, . \, \frac{ \vec{u} }{ ||\vec{u}|| } \right)^2 \; + \; \left( \vec{\gamma} \, . \, \frac{ \vec{v} }{ ||\vec{v}|| } \right)^2 \; \right]
\end{equation}
The second term of the previous equation is the reduced statistics used in the following:

\begin{equation}
\Lambda_{\text{reduced}} \; = \; \left[ \; \left( \vec{\gamma} \, . \, \frac{ \vec{u} }{ ||\vec{u}|| } \right)^2 \; + \; \left( \vec{\gamma} \, . \, \frac{ \vec{v} }{ ||\vec{v}|| } \right)^2 \; \right]
\end{equation}

From the definitions of the vectors $\vec{\gamma}$, $\vec{u}$ and $\vec{v}$, it clearly follows that the two Gaussian variables $\vec{\gamma} . \frac{\vec{u}}{||\vec{u}||}$ and $\vec{\gamma} . \frac{\vec{v}}{||\vec{v}||}$ are uncorrelated. Indeed, the distribution of $\Lambda_{\text{reduced}}$ is close to a $\chi^2$ variable with two degrees of freedom, independently of the source location in the sky and of the particular network considered.

\subsection{Coherent data analysis ROC}

\begin{figure}
\centerline{\epsfig{file=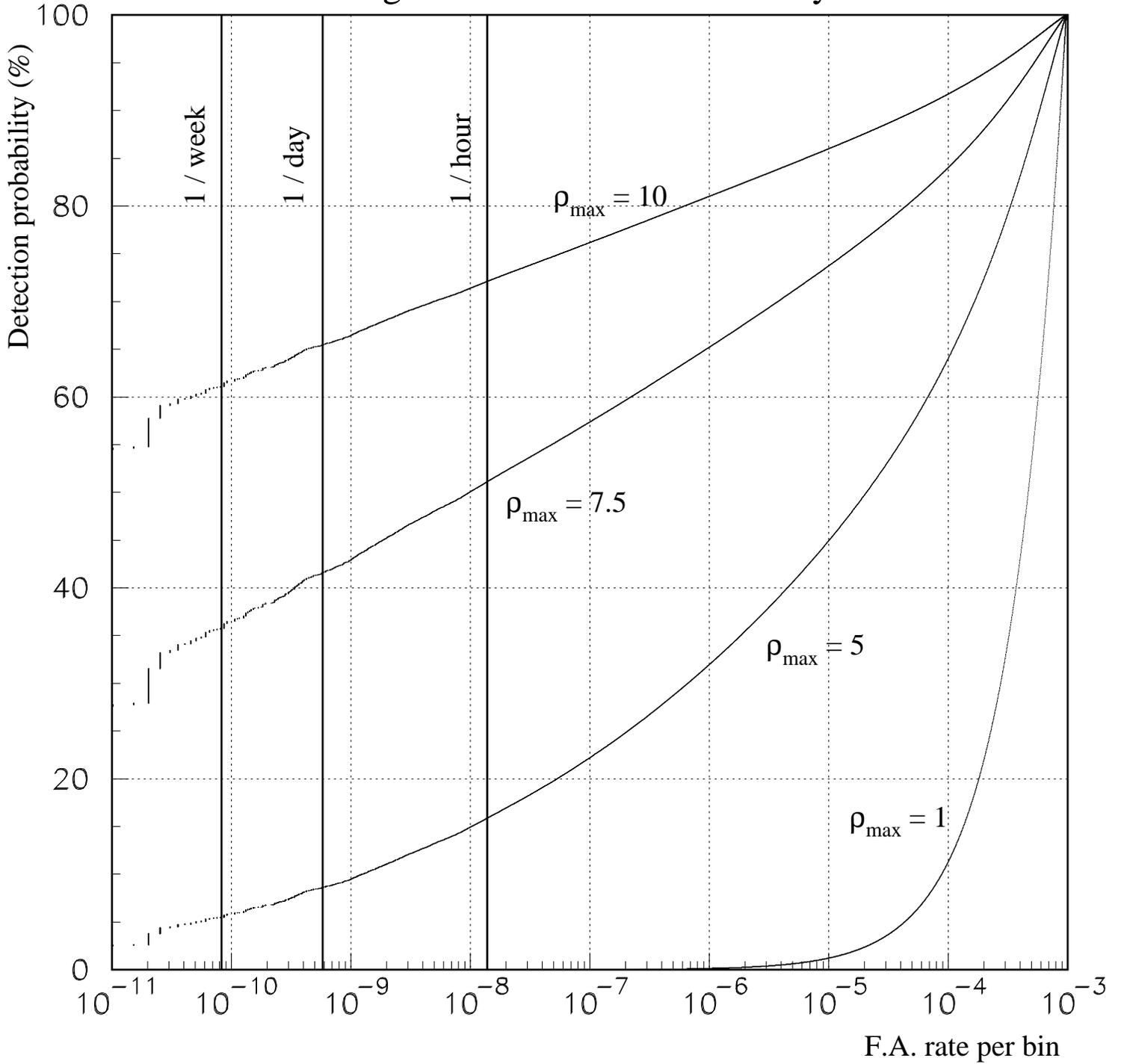,width=20cm}}
\caption{ROC for the coherent search of GW burst signals in the three-interferometer network Virgo-LIGO, assuming the source sky location to be known. The curves have been computed for four different values of the optimal SNR: $\rho_{\text{max}}=1$, 5, 7.5 and 10 respectively.}
\label{fig:roc_likelihood_vhl}
\end{figure}

\begin{figure}
\centerline{\epsfig{file=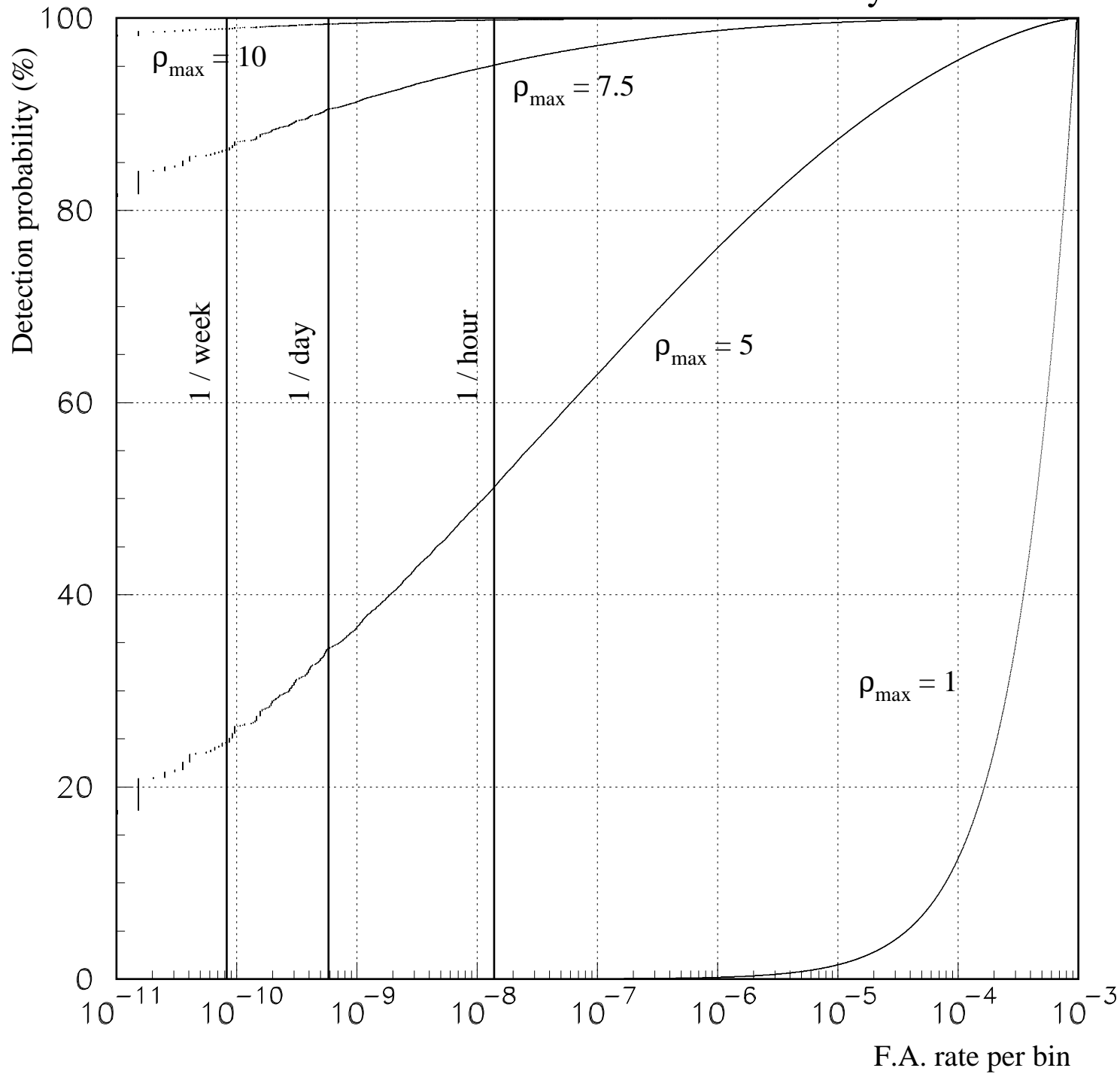,width=20cm}}
\caption{ROC for the coherent search of GW burst signals in the full network of interferometers including the 6 currently existing projects in the world. For $\rho_{\text{max}}=10$, the detection efficiency is higher than 95\% in the whole range of $\tau$ considered and $\epsilon > 80\%$ for the case $\rho_{\text{max}}=7.5$, assuming the source location to be known.}
\label{fig:roc_likelihood_full}
\end{figure}

Figures \ref{fig:roc_likelihood_vhl} and \ref{fig:roc_likelihood_full} present two examples of coherent ROC, for the three detector network Virgo-LIGO and for the full set of six interferometers. For the two networks considered here, the coherent analysis ROC are clearly above all coincidence ROC. This is mostly due to the more complete management of data in the coherent method case. Moreover, comparing with the loose coincidence case, the coherent approach benefits in addition from the fact that the source location is known. Section \ref{section:comparison} summarizes the comparison of both network data analysis approaches.

One can also note that going from three to six interferometers strongly increases the detection probabilities; the differences are more significant than for the coincidence case. Indeed, for $\rho_{\text{max}}=10$, the efficiency remains higher than 97\% in the whole range of false alarm rates per bin; for $\rho_{\text{max}}=7.5$, the detection efficiency is at least 80\%. For $\tau=1$/hour and $\rho_{\text{max}}=5$, one has still $\epsilon=50\%$.

\subsection{Coherent data analysis timing accuracy}

\begin{figure}
\centerline{\epsfig{file=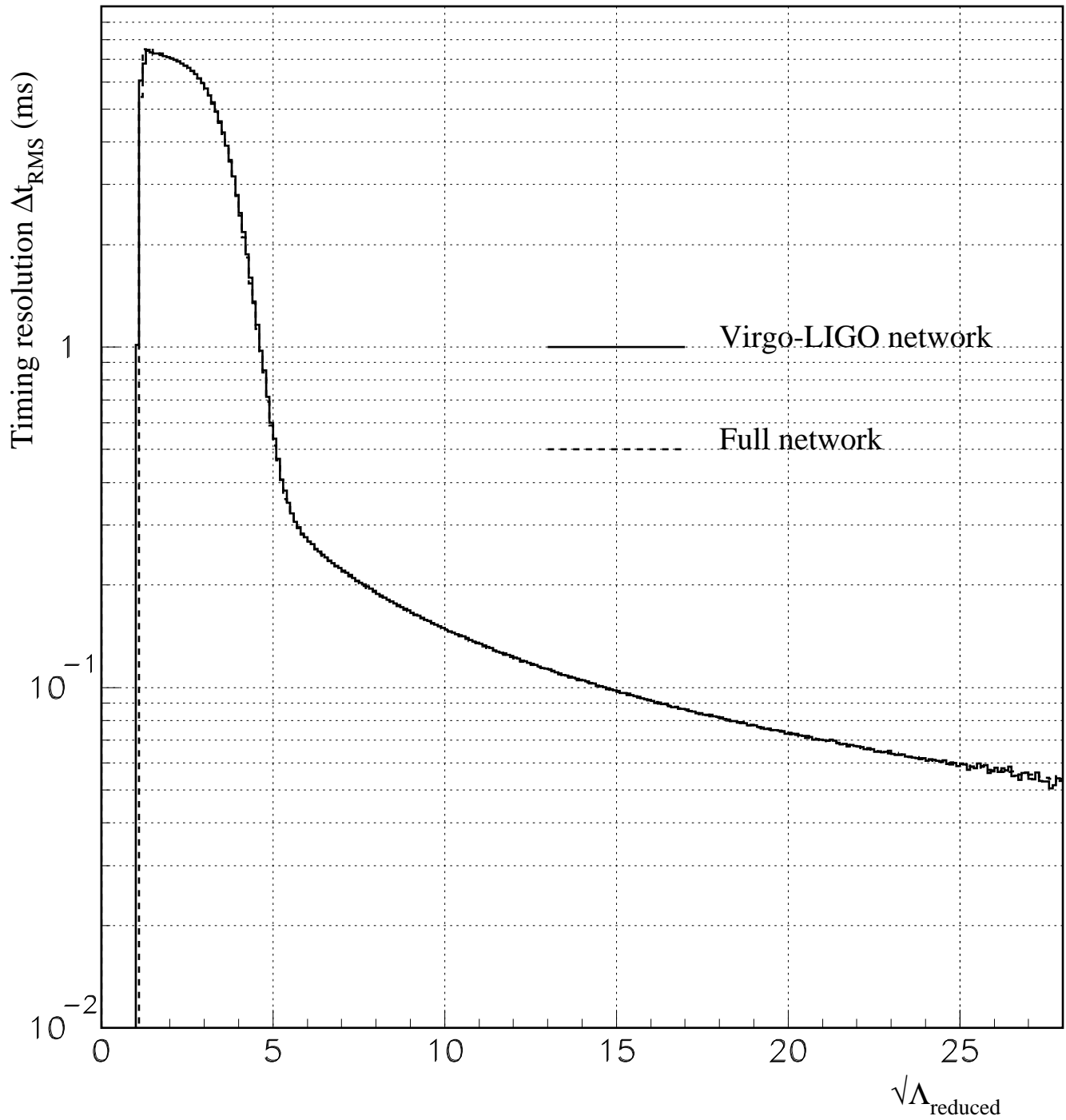,width=20cm}}
\caption{Timing resolution of the coherent analysis showing the evolution of the timing error $\Delta t_{\text{RMS}}$ (in ms) as a function of the square root of the coherent statistics $\left(\Lambda_{\text{reduced}}\right)^{1/2}$. The two curves -- for the Virgo-LIGO network and for the full set of interferometers -- are identical, as expected from the network-independent statistics $\Lambda_{\text{reduced}}$.}
\label{fig:coherent_timing_efficiency}
\end{figure}

As for the single detector case, the timing accuracy of the coherent method can be easily studied -- estimating the timing accuracy for coincidences is not as straightforward. Figure \ref{fig:coherent_timing_efficiency} shows the evolution of the timing error $\Delta t_{\text{RMS}}$ as a function of $\left(\Lambda_{\text{reduced}}\right)^{1/2}$ for the two examples of networks considered in this section: Virgo-LIGO and the full set of six interferometers. As for the single detector case, the precision goes well below the signal half-width $\omega$ -- taken equal to 1~ms here. Taking the square-root of the (quadratic) coherent statistics is mandatory in order to have a quantity scaling with the optimal SNR $\rho_{\text{max}}$. Due to the 'universality' of $\Lambda_{\text{reduced}}$, the two curves presented on the plot overlap perfectly.

One can also try to connect $\left(\Lambda_{\text{reduced}}\right)^{1/2}$ and $\rho_{\text{max}}$, at least in average. For $\rho_{\text{max}} \ge 3$, linear fits give:

\begin{eqnarray}
\nonumber
\overline{ \left(\Lambda_{\text{reduced}}\right)^{1/2} } &\approx& 0.67 \, \times \, \rho_{\text{max}} \; + \; 0.89 \;\;\;\; \text{for the Virgo-LIGO network} \\ 
\nonumber
\overline{ \left(\Lambda_{\text{reduced}}\right)^{1/2} } &\approx& 1.06 \, \times \, \rho_{\text{max}} \; + \; 0.35 \;\;\;\; \text{for the six-interferometer network}
\end{eqnarray}
Inverting these equations allows one to roughly link $\left(\Lambda_{\text{reduced}}\right)^{1/2}$ to a value of the optimal SNR $\rho_{\text{max}}$. Of course, the larger the network, the higher the mean value of $\left(\Lambda_{\text{reduced}}\right)^{1/2}$ at fixed $\rho_{\text{max}}$. 

The proper way to estimate the timing accuracy improvement provided by a network coherent analysis with respect to the single interferometer case is to compare the timing error RMS for a given optimal SNR $\rho_{\text{max}}$. Table~III below shows the timing performances of the different configurations for three values of $\rho_{\text{max}}$: 5, 7.5 and 10 respectively. Only events exceeding the threshold tuned at a false alarm rate of 1/hour are included in the computation.

\begin{center}
\begin{tabular}{|c|c|c|c|}
\hline $\rho_{\text{max}}$ & Single detector & Virgo-LIGO network & Full network \\
& & (coherent analysis) & (coherent analysis) \\
\hline 5 & 0.44 ms & 0.29 ms & 0.25 ms \\
\hline 7.5 & 0.29 ms & 0.23 ms & 0.18 ms \\
\hline 10 & 0.24 ms & 0.19 ms & 0.14 ms \\
\hline
\end{tabular}
\end{center}
\vskip -0.2truecm
\centerline{Table~III: Timing performance comparison for different values of $\rho_{\text{max}}$.}

As expected, the coherent analysis improves also the timing precision, especially at low optimal SNR. GW events do not only trigger more often; they are also more precisely located. Finally, Table~IV and V show how the coherent timing performances at given $\rho_{\text{max}}$ evolve when the false alarm rate is reduced. As the thresholds increase, the quality of the selected sample improves; yet, the precision in locating the GW signal peaks does not change significantly.

\begin{center}
\begin{tabular}{|c|c|c|c|}
\hline False alarm rate & $\rho_{\text{max}}=5$ & $\rho_{\text{max}}=7.5$ & $\rho_{\text{max}}=10$ \\
\hline 1/hour & 0.29 ms & 0.23 ms & 0.19 ms \\
\hline 1/day & 0.25 ms & 0.21 ms & 0.18 ms \\
\hline 1/week & 0.24 ms & 0.20 ms & 0.17 ms \\
\hline
\end{tabular}
\end{center}
\vskip -0.2truecm
\centerline{Table~IV: Coherent timing accuracy for decreasing false alarm rates in the Virgo-LIGO network.} 

\begin{center}
\begin{tabular}{|c|c|c|c|}
\hline False alarm rate & $\rho_{\text{max}}=5$ & $\rho_{\text{max}}=7.5$ & $\rho_{\text{max}}=10$ \\
\hline 1/hour & 0.25 ms & 0.18 ms & 0.14 ms \\
\hline 1/day & 0.24 ms & 0.18 ms & 0.14 ms \\
\hline 1/week & 0.23 ms & 0.18 ms & 0.14 ms \\
\hline
\end{tabular}
\end{center}
\vskip -0.2truecm
\centerline{Table~V: Coherent timing accuracy for decreasing false alarm rates in the full network.} 

\section{Matched filtering of the celestial sphere for a coherent analysis with a network of interferometers}
\label{section:sky_tiling}

The last step of the coherent search -- when the source location is a priori unknown -- consists in estimating the number of filters $\mathfrak{N}$ needed for the sky coverage. To do this, the most efficient way is to use the method first defined in Ref. \cite{Owen} for the in-spiral binary case, and then extended for coherent analysis of Newtonian chirp binary signals in a network up to three interferometers \cite{PDB}. Here, we still use a Gaussian peak of width $\omega$ as 'generic' GW burst signals.

The main difference with the matched filtering case is that there is no more symmetry between the interferometer data and the template. Quantifying the separation between two close filters is thus not easy. For instance, let us consider the case of the polarization angle $\psi$; as shown in section \ref{analysis:deriv_stat}, there exists an analytical way to maximize $\Lambda$ over $\psi_{\text{template}}$ while $\psi_{\text{signal}}$ remains 'hidden' in the noisy data. The solution proposed by Ref. \cite{PDB} is to choose some values for $\psi_{\text{signal}}$ -- and also for the binary orbit inclination in that case -- and to estimate the number of templates $\mathfrak{N}$ for these different configurations. Numerically, it is found that $\mathfrak{N}$ does not change by more than a factor 3 in the range of parameters tested. In this paper, a different path is followed: the logarithm of the likelihood ratio is first averaged over $\psi$, which allows one to focus only on the sky angular dependence of the beam pattern function.

Before presenting the calculation one can remark that, as first pointed out in \cite{PDB}, the loss in SNR caused by a mismatch in the source direction is mainly due to the corresponding wrong time delays.

\subsection{Ambiguity function and metric in the celestial coordinates}

Assuming a mismatch $\delta t^i$ between the Gaussian peak template and the GW signal -- both of characteristic width $\omega$ --, a straightforward calculation of the correlation gives:

\begin{equation}
\langle s^i \, | \, x^i \rangle \; = \; K^2 \; \left( F^i \right)^2 \; \exp \left[ - \left( \frac{ \delta t^i }{ 2\, \omega } \right)^2 \right]
\label{eq:reduction}
\end{equation}
The exponential term reduces the maximal correlation and the signal width $\omega$ provides a timescale with which the error $\delta t^i$ is compared. Of course, Eq. (\ref{eq:reduction}) would be meaningless in the case of a single detector, as a simple opposite time-shift of the template would allow one to recover the full SNR. In coherent analysis, time shifts cannot be optimized separately for each interferometer; therefore, a wrong match of the detection leads to unavoidable losses in SNR. The logarithm of the likelihood ratio averaged on the polarization angle is thus equal to:

\begin{equation}
\ln\lambda \; = \; K^2 \; \sum_{i=1}^{P} \; \left( \varphi^i \right)^2 \; \left[ \, \exp \left[ - \left( \frac{ \delta t^i }{ 2\, \omega } \right)^2 \right] \; - \; \frac{ 1 }{ 2 } \, \right] \;\;\; \text{with} \;\;\; \varphi^i \; = \; \frac{ \overline{F}^i }{ \sigma^i }
\label{eq:LR_averaged}
\end{equation}

To 'transform' $\ln\lambda$ into an ambiguity function giving the relative mean loss in SNR due to the direction mismatch, one chooses $K$ such as $\ln\lambda|_{(\delta t^i=0)}=1$, which is achieved with:

\begin{equation}
\nonumber
K \; = \; \sqrt{ \frac{ 2 }{ \sum_{i=1}^{P} \; \left( \varphi^i \right)^2 } }
\end{equation}

A Taylor expansion $\mathbb{A}$ around $(\delta t^i=0)_{i=1,...,P}$ at the second order gives the quadratic approximation of the ambiguity function, assumed to be valid provided that the allowed losses of SNR remain small.

\begin{equation}
{\mathbb{A}} \; = \; 1 \; - \; \frac{ 1 }{ 4 \, \omega^2 } \; \frac{ \sum_{i=1}^{P} \; \left( \varphi^i \right)^2 \; \left( \delta t^i \right)^2 }{ \varphi^2 } \;\;\; \text{with} \;\;\; \varphi^2 \; = \; \sum_{i=1}^{P} \; \left( \varphi^i \right)^2
\label{eq:A}
\end{equation}

Then, one has to replace the $\delta t^i$ by their expressions in term of the two angular variables locating the source direction in the sky: the right ascension $\alpha$ and the sine of the declination $X=\sin\delta$. Let $\Omega$ represent the center of Earth and $\vec{n}$ be the unit vector radiating from it in the source direction. One has

\begin{equation}
\delta t^i \; = \; - \frac{ 1 }{ \light } \; \overrightarrow{ \delta n } \; . \; \overrightarrow{ \Omega D^i } 
\nonumber
\end{equation}
with $\light$ being the speed of light and $\overrightarrow{ \delta n }$ the error in the direction of the source location -- note that from Eq. (\ref{eq:LR_averaged}) the sign convention of $\delta t^i$ does not matter. The computation of $\overrightarrow{ \delta n }$ is straightforward:

\begin{equation}
\overrightarrow{ \delta n } \; = \; 
\begin{pmatrix}
- \sqrt{ 1 \, - \, X^2 } \; \sin\alpha & - \frac{ X \, \cos\alpha }{ \sqrt{ 1 \, - \, X^2 } } \\
\sqrt{ 1 \, - \, X^2 } \; \cos\alpha & - \frac{ X \, \sin\alpha }{ \sqrt{ 1 \, - \, X^2 } } \\
0 & 1
\end{pmatrix}
\begin{pmatrix}
d\alpha \\
dX
\end{pmatrix}
\nonumber
\end{equation}
The $3 \times 2$ matrix appearing in the previous equation will be designed as $\mathbb{M}$ in the following. In order to shorten the expressions appearing in the metric calculation, one can also introduce a $3 \times 3$ matrix $\Gamma$ defined as follows:

\begin{equation}
\Gamma_{kl} \; = \; \sum_{i=1}^{P} \; \left( \varphi^i \, \kappa_k^i \right ) \; \times \;  \left( \varphi^i \, \kappa_l^i \right ) \;\;\; \text{with} \;\;\;  \overrightarrow{ \Omega D^i } \; = \; 
\begin{pmatrix}
\kappa_1^i \\
\kappa_2^i \\
\kappa_3^i \\
\end{pmatrix}
\nonumber
\end{equation}
The $\Gamma$ matrix contains all the network characteristics. Equation (\ref{eq:A}) can thus be rewritten:

\begin{equation}
{\mathbb{A}} = 1 \; - \; \frac{ 1 }{ 4 \, \omega^2 \, \light^2 \, \varphi^2 } \; ^t\left(\overrightarrow{ \delta n }\right) \; . \; \Gamma \; . \;  \overrightarrow{ \delta n }
\; = 1 \; - \; \frac{ 1 }{ 4 \, \omega^2 \, \light^2 \varphi^2 } \; 
\begin{pmatrix}
d\alpha & dX
\end{pmatrix}
\; . \; \underbrace{ \left( ^t {\mathbb{M}} \; \Gamma \; {\mathbb{M}} \right) }_{ = \; G } \; . \; 
\begin{pmatrix}
d\alpha \\ dX
\end{pmatrix}
\label{eq:A_G}
\end{equation}
The tiling metric $g$ is thus:

\begin{equation}
g \; = \; \frac{ G }{ 4 \, \omega^2 \, \light^2 \, \varphi^2 }
\nonumber
\end{equation}
The exact expressions for the coefficients of the (symmetrical) $2 \times 2$ matrix $G$ are given in Appendix \ref{appendix:metric_matrix}, so as its determinant $\Delta_G$.

\subsection{Number of templates needed for the various network configurations}

Calculating the metric allows one to estimate the number of templates ${\mathfrak{N}}_{2D}$ needed to cover the whole sky for a particular network of interferometers, given the maximal allowed loss in SNR. This last quantity is usually written as $1 - MM$ where $MM$ is the 'Minimal Match' \cite{Owen} -- a conventional value is $MM = 97\%$.

Still following Ref. \cite{Owen}, ${\mathfrak{N}}_{2D}$ is computed by integrating over the sky the square root of the metric determinant, multiplied by a scaling factor depending on the minimal match and on the parameter space $\{(\alpha,X)\}$. One gets finally

\begin{equation}
{\mathfrak{N}}_{2D} \; \sim \; \frac{ 1 }{ 8 \, \omega^2 \, \light^2 \, ( 1 \, - \, MM ) } \; \int_{ [-\pi;\pi] \times [-1;1] } \; \left[ \; \frac{ \sqrt{ \Delta_G } }{ \, \varphi^2 } \; \right] \; d\alpha \, dX 
\label{eq:N}
\end{equation}

From this formula, one can note that the longer the signal, the smaller the number of templates. This last feature is due to the particular burst shape chosen in the paper and cannot be generalized to any GW signal (indeed pure sines behave in an exactly opposite way as the larger their number of cycles, the more they need to be accurately tracked in noisy data). Appendix \ref{appendix:list} gives an estimation of $\mathfrak{N}$ for the different networks of existing interferometers.

One can see that the smaller the network, the more the value of $\mathfrak{N}$ depends on the particular configuration. This is particularly true for the case of $P=2$ detectors for which there is a factor higher than six between the extreme values. In this case, the number of templates does not only depend on the light-distance between the two interferometers but also on their respective orientations. For larger networks, the results are closer: the exact locations of the detectors appear less important, they look like more 'randomly' spaced on Earth. For the set of 6 interferometers, one has ${\mathfrak{N}} \sim 5320$.

\subsection{Extending the space parameter}

One can also assume that the exact width of the GW signal is not known and thus that $\omega$ is another parameter of the search. It is easy to check that the only change in Eq. (\ref{eq:A}) is the apparition of a new term reducing in addition the ambiguity function:

\begin{equation}
\nonumber
- \; \frac{ 1 }{ 4 } \; \left( \frac{ \delta \omega }{ \omega } \right)^2
\end{equation}
where $\delta \omega$ is the error on the Gaussian peak width. As the width and the angular parameters are decoupled, estimating $\mathfrak{N}$ is straightforward:

\begin{equation}
{\mathfrak{N}}_{3D} \; \sim \; \frac{ 3 \, \sqrt{3} }{ 64 \, \, \light^2 \, ( 1 \, - \, MM )^{3/2} } \; \left( \; \int_{ [-\pi;\pi] \times [-1;1] } \; \left[ \; \frac{ \sqrt{ \Delta_G } }{ \, \varphi^2 } \; \right] \; d\alpha \, dX \; \right) \; \left( \; \int_{[\omega_{\text{min}};\omega_{\text{max}}]} \; \frac{ d\omega }{ \omega^3 } \; \right)
\label{eq:N_3D}
\end{equation}

To measure the 'template cost' due to the addition of the third free parameter $\omega$, one can for instance compute the ratio between the '3D' number of filters needed to fill both the $\omega$-range $[\omega_{\text{min}};\omega_{\text{max}}]$ and the corresponding '2D' number for $\omega=\omega_{\text{min}}$ fixed. As seen from Equations (\ref{eq:N}) and (\ref{eq:N_3D}), this ratio does not depend on the network as the angular integrals simplify.

\begin{equation}
\frac{ {\mathfrak{N}}_{3D} \left[ \omega_{\text{min}};\omega_{\text{max}} \right] }{ {\mathfrak{N}}_{2D} \left( \omega_{\text{min}} \right) } \; = \; \frac{ 3 \, \sqrt{ 3 } }{ 16 \, \sqrt{ 1 \, - \, MM } } \; \left[ 1 \, - \, \left( \frac{ \omega_{\text{min}} }{ \omega_{\text{max}} } \right)^2 \right]
\end{equation}

For $MM=97\%$, the numerical factor in front of the brackets is equal to 1.88 and this asymptotic value is quickly reached when the ratio $\omega_{\text{min}} / \omega_{\text{max}}$ decreases. The number of templates only doubles when one goes from two to three parameters; thus, covering coherently the celestial sphere for a burst search over a wide range of durations is not too expensive. This number has to be compared with the overestimated value of ${\mathfrak{N}}_{3D}$ computed by multiplying ${\mathfrak{N}}_{2D}$ by the number of filters ${\mathfrak{N}}_{\omega}$ needed to cover the one-dimensional parameter space $[ \omega_{\text{min}};\omega_{\text{max}} ]$. As ${\mathfrak{N}}_{\omega}=12$ for the numerical data considered in this section \cite{arnaud_these}, the saving in template number -- and thus in CPU time -- is at least a factor 6.

\section{Comparison of coincidences and coherent data analysis methods}
\label{section:comparison}

As previously stated, comparing the ROC for coherent and coincident analysis leads to the clear conclusion that the former approach shows better performances. Indeed, methods based on a coherent use of the various datasets must give better results than coincidences, as the merging of informations provided by the different interferometers is more complete than a simple binary test (absence or presence of the signal in a given detector).

A very strong assumption made for the study of the coherent analysis is that the source location is known, whereas no such hypothesis is necessary for loose coincidence detections. A priori, this additional information could be the main origin of the performance differences between the two network data analysis methods. Yet, the studies performed in this paper do not confirm this hypothesis of the dominant improvement factor. Indeed, tight coincidences have also been studied. In this case, the source location is assumed to be known, as for the coherent analysis, and so ROC are directly comparable. Yet, the performance gap between the two network algorithms remains wide.

Requiring the knowledge of the source location for coherent filtering has also another consequence: to cover the full sky, many templates must be used in parallel. Therefore, the meaningful quantity is no more the false alarm rate per bin $\tau_{\text{single}}$ of a given coherent filter, but rather the global false rate $\tau_{\text{global}}$, computed by taking into account the whole set of templates. As the parameter space grid is thin, false alarms between close filters are certainly correlated: if one filter triggers, some templates corresponding to neighbor locations should also exceed the threshold. Computing the correlation level is a complete work by itself; thus, in this article, we only estimated it roughly with a toy Monte-Carlo cheating on the precise location of the templates. Assuming 1 false alarm per hour and per template, the fraction of filters triggering simultaneously is $\kappa_{\text{correl}} \approx 7 \%$ for the Virgo-LIGO network and $\kappa_{\text{correl}} \approx 0.5 \%$ for the full set of 6 interferometers. As

\begin{equation}
\tau_{\text{global}} \; \sim \; \tau_{\text{single}} \; \times \; \kappa_{\text{correl}} \; \times \; {\mathfrak{N}}
\end{equation}
with $\mathfrak{N}$ being the number of templates computed in the previous section. From the numerical results given in Appendix \ref{appendix:list}, one can deduce that with a minimal match $MM=97\%$ one has

\begin{equation}
\tau_{\text{global}} \; \approx \;
\begin{cases}
350 \; \times \; \tau_{\text{single}} \;\;\; \text{for the Virgo-LIGO network} \\
25 \; \times \; \tau_{\text{single}} \;\;\; \text{for the full network}
\end{cases}
\end{equation}
as the template number is ${\mathfrak{N}} \sim 5000$ in both cases.

Comparing Figures \ref{fig:roc_single} to \ref{fig:roc_coincid_full} on the one hand and Figures \ref{fig:roc_likelihood_vhl}-\ref{fig:roc_likelihood_full} on the other hand, clearly shows that even if the horizontal-axis of the coherent ROC are shifted by these values on the right, the corresponding detection probabilities remain clearly higher than for the coincidence methods.

Indeed, Table~VI summarizes the coincidence and (rescaled) coherent analysis efficiencies\footnote{The values tagged with an (*) in Table~VI have been estimated by prolongating the ROC beyond the range of false alarm rates achieved by the numerical simulations.} for different false alarm rates. Three scenarii are compared:

\begin{itemize}
\item $\tau_{\text{global}} = \tau_{\text{single}}$: no correlation between templates;
\item $\kappa_{\text{correl}}$ equal to the estimations presented above;
\item the worst (and unlikely) case, $\kappa_{\text{correl}} = 1$: maximum correlation.
\end{itemize}

\begin{center}
\begin{tabular}{|c|c|c|c|}
\hline False alarm rate $\tau_{\text{norm}}$ & 1/week & 1/day & 1/hour \\
\hline Twofold loose coincidence & 34\% & 37\% & 44\% \\
\hline Twofold tight coincidence & 37\% & 41\% & 47\% \\
\hline Coherent analysis (no correlation) & 61\% & 65\% & 72\% \\
\hline Coherent analysis ($\kappa_{\text{correl}} = 7 \%$) & 46\% (*) & 53\% (*) & 60\% \\
\hline Coherent analysis ($\kappa_{\text{correl}} = 1$) & 43\% (*) & 48\% (*) & 54\% (*) \\
\hline
\end{tabular}
\vskip 0.2truecm
Virgo-LIGO network \\
\vskip 0.2truecm
\begin{tabular}{|c|c|c|c|}
\hline False alarm rate $\tau_{\text{norm}}$ & 1/week & 1/day & 1/hour \\
\hline Twofold loose coincidence & 69\% & 74\% & 83\% \\
\hline Twofold tight coincidence & 76\% & 82\% & 87\% \\
\hline Threefold loose coincidence & 52\% & 57\% & 65\% \\
\hline Threefold tight coincidence & 61\% & 64\% & 72\% \\
\hline Coherent analysis (no correlation) & 98\% & 99\% & 100\% \\
\hline Coherent analysis ($\kappa_{\text{correl}} = 0.5 \%$) & 98.1\% (*) & 98.6\% & 99.3\% \\
\hline Coherent analysis ($\kappa_{\text{correl}} = 1$) & 96.0\% (*) & 96.7\% (*) & 97.6\% (*) \\
\hline
\end{tabular}
\vskip 0.2truecm
Full network \\
\end{center}
\vskip -0.2truecm
\centerline{Table~VI: Comparison between coincidence and (rescaled) coherent analysis detection efficiencies at various false alarm rates. }

Finally, one can note that keeping $\kappa_{\text{correl}}$ constant when $\tau_{\text{norm}}$ decreases leads to an overestimation of the template correlations at smaller false alarm rates: the higher the threshold, the smaller the probability to have again the filter outputs triggering when the datasets are shifted one with respect to the other. Therefore, coherently analyzing data coming from the different detectors increases significantly in all cases the detection potential of interferometer networks. Moreover, the number of templates involved in such searches appears low enough to make these analyzes feasible with a small CPU farm.

The main reason why coherent analysis appears so successful is certainly its capability to sum the signal contributions from the different interferometers regardless whether they individually trigger. A coherent detection can perfectly originate from outputs distributed in such a way that none of them triggers on coincidence strategies for thresholds adjusted to the same false alarm rate! On the other hand, coincidences always loose a significant fraction of the available information, which becomes more and more important as the network size increases: the larger the network, the more likely that a GW signal strong enough is above the background noise in some of these detectors.

\section{Conclusion}

Coincidence and coherent data analysis methods in networks of interferometric GW detectors are compared in this article thanks to a network model which allows one to compare these two approaches quantitatively, through ROC curves. Indeed, these graphs summarize well the behavior of a particular algorithm for a given GW signal over a wide range of false alarm rates. First, coincidence methods are studied in different networks from two to six interferometers. To select events, two different compatibility criteria are defined. The first one, the loose test, does not require any assumption on the source location in the sky; therefore, using it allows one to search events in the whole celestial sphere, however with limited efficiency.

From this study, it clearly appears that searching GW bursts in a single detector is not efficient at all. 
For what concerns two detector networks, the LIGO 4-km pair is the most efficient, due to their relative closeness and especially to their 'parallel' orientation. Yet, detection efficiencies remain limited for such reduced networks. Therefore, larger sets of detectors have to be considered. Adding Virgo to the two LIGO detectors shows a significant enhancement of the twofold detection efficiency. But higher-fold coincidences remain improbable, unless the network size increases significantly. So, the goal of a worldwide coincidence analysis should be to include as much interferometers as possible -- indeed, six would not be too much! -- in the network.

A complementary study of the full LIGO network including the two 4-km interferometers and the 2-km detector in Hanford shows that differences in the sensitivity of network components have important consequences on the network performances: the factor two difference in the 2-km interferometer sensitivity is more important than the false alarm rate reduction due to the close location of the two Hanford detectors. Therefore, an efficient network should contain interferometers with sensitivities as close as possible. Conversely, adding less efficient instruments to a network is not productive.

Finally, few ROC about tight coincidences are presented. As the source location is known, the compatibility test is more constraining, leading thus to an improvement of the ROC performances. Yet, the gain is small as coincidence analyses are limited by the loss of information due to the binary diagnostic made in each interferometer of the network. On the other hand, coherent analyses benefit from all detector outputs without a priori on the presence/absence of a GW signal in the data, and are so much more powerful.

The difference in performances between coincidence and coherent analysis is significant in both networks considered here: Virgo-LIGO and the full network of six detectors. Coherent detection efficiencies remain very large even at small false alarm rate for $\rho_{\text{max}}=10$ -- more than 95\% at $\tau_{\text{norm}}=1/$week in the six detector network! On the other hand, weaker signals (say $\rho_{\text{max}}=5$ or below) are not well seen: the final sensitivities of network components will be critical. Finally, the timing accuracy of coherent data analysis method is also considered; as expected, it is in average better than the single interferometer case, and improves with the network size.

Another point worth being mentioned about coherent analysis methods is that they could be used even when the waveform is not accurately known, like for GW bursts in general. Indeed, the only assumption made in this paper is that the filtering algorithm is linear -- cf. Eq. (\ref{eq:filtering}). For instance, one could also use some robust and efficient filters \cite{arnaud_burst_1,hello_timing} which only depend on a single parameter, the analysis window size.

The price to pay for the high performances of the coherent data analysis method is its complexity, especially with respect to the loose coincidence approach. But, at least in the case of burst signals, this does not appear to be a strong limitation: the number of templates needed to scan accurately the whole sky is quite small -- at most a few thousands --, even including the signal width. In addition, the correlation between the templates is estimated to be below a few percent or even less. Therefore, the loss in performances induced by the increase of the global coherent false alarm rate with respect to the single template case is limited: coherent methods are better than coincidence searches.

So, the main conclusion of this study is that one should not limit collaborative data analysis to the exchange of single interferometer events, especially for GW bursts. Otherwise, a large fraction of detection efficiency will be lost, which may be crucial for rare sources, like e.g. close supernovae. Using the full set of available data for GW signal search -- in the largest possible network -- should be an important goal of the worldwide GW data analysis community, at least in a mid-term perspective.

\appendix

\section{Coefficients and determinant of the metric matrix $G$}
\label{appendix:metric_matrix}

With the notations defined in the core of the paper, the coefficients of the $2 \times 2$ symmetrical matrix $G$ are:

\begin{eqnarray}
\nonumber
G_{11} &=& ( 1 \; - \; X^2 ) \; \left[ \; \Gamma_{11} \, \sin^2\alpha \; - \Gamma_{12} \, \sin(2\alpha) \; + \; \Gamma_{22} \, \cos^2\alpha \; \right] \\ 
\nonumber
G_{12} &=& G_{21} \; = \; X \; \left[ \; \frac{ ( \Gamma_{11} \, - \, \Gamma_{12} ) \, \sin2\alpha }{ 2 } \; - \; \Gamma_{12} \, \cos2\alpha \; \right] \; + \; \sqrt{ 1 \, - \, X^2 } \; \left( \; \Gamma_{23}\cos\alpha \, - \, \Gamma_{13} \, \sin\alpha \; \right) \\
\nonumber
G_{22} &=& \frac{ X^2 }{ 1 \, - \, X^2 } \; \left( \; \Gamma_{11} \, \cos^2\alpha \; + \; \Gamma_{22} \, \sin^2\alpha \; + \; \Gamma_{12} \, \sin2\alpha \; \right) \; - \; \frac{ 2 X }{ \sqrt{ 1 \, - \, X^2 } } \; \left( \Gamma_{13} \, \cos\alpha \; + \; \Gamma_{23} \, \sin\alpha \right)
\end{eqnarray}

To estimate the number of templates $\mathfrak{N}$, one needs to compute the determinant of $G$. Extensive calculations give:

\begin{eqnarray}
\nonumber
\Delta_G &=& \left( \; \Gamma_{11} \, \Gamma_{22} \; - \; \Gamma_{12}^2 \; \right) \; X^2 \\
\nonumber
&\;& - \; 2 \, X \, \sqrt{ 1 \, - \, X^2 } \; \left[ \; \left( \Gamma_{22} \, \Gamma_{13} \; - \; \Gamma_{12} \, \Gamma_{23} \right) \, \cos\alpha \; + \; \left( \Gamma_{11} \, \Gamma_{23} \; - \; \Gamma_{12} \, \Gamma_{13} \right) \, \sin\alpha \; \right] \\
\nonumber
&\;& + \; ( 1 \, - \, X^2 ) \; \left[ \; \left( \Gamma_{22} \, \Gamma_{33} \, - \, \Gamma_{23}^2 \right) \, \cos^2\alpha \; + \; \left( \Gamma_{11} \, \Gamma_{33} \, - \, \Gamma_{13}^2 \right) \, \sin^2\alpha \; + \; \left( \Gamma_{13} \, \Gamma_{23} \, - \, \Gamma_{12} \, \Gamma_{33} \right) \sin2\alpha \; \right]
\end{eqnarray}

As shown by the previous formula, $\Delta_G$ is never singular, even in the directions $X = \pm 1$.

\section{List of the numbers of templates for the different configurations of detectors}
\label{appendix:list}

This appendix gives the estimated number of templates needed to cover the whole sky for each possible network of interferometers: from 2 detectors to the whole set of 6 antennas. These numbers are computed with $\omega=1$ ms and $MM = 97\%$ and thus must be properly rescaled for different choices of these two coefficients -- see Eq. (\ref{eq:N}).

In the following tables compiling the results of the calculations, the interferometers are simply designed by a single letter: Virgo (V), LIGO Hanford (H) and LIGO Livingston (L), GEO600 (G), TAMA300 (T) and ACIGA (A).

\begin{center}
\begin{tabular}{||c|c||c|c||c|c||}
\hline Configuration & $\mathfrak{N}$ & Configuration & $\mathfrak{N}$ & Configuration & $\mathfrak{N}$ \\
\hline V-H & 4980 & H-L & 2580 & L-T & 5070 \\
\hline V-L & 4770 & H-G & 4680 & L-A & 2040 \\
\hline V-G & 800 & H-T & 4870 & G-T & 5060 \\
\hline V-T & 5060 & H-A & 3590 & G-A & 4180 \\
\hline V-A & 4190 & L-G & 5100 & T-A & 4710 \\
\hline
\end{tabular}
\end{center}
\vskip -0.2truecm
\centerline{2 interferometer networks}

\begin{center}
\begin{tabular}{||c|c||c|c||c|c||c|c||}
\hline Configuration & $\mathfrak{N}$ & Configuration & $\mathfrak{N}$ & Configuration & $\mathfrak{N}$ & Configuration & $\mathfrak{N}$ \\
\hline V-H-L & 5020 & V-L-T & 5060 & H-L-G & 5000 & H-T-A & 5430 \\
\hline V-H-G & 4880 & V-L-A & 4800 & H-L-T & 5200 & L-G-T & 5220 \\
\hline V-H-T & 5130 & V-G-T & 5050 & H-L-A & 3460 & L-G-A & 4820 \\
\hline V-H-A & 5050 & V-G-A & 4260 & H-G-T & 5090 & L-T-A & 5080 \\
\hline V-L-G & 4910 & V-T-A & 5150 & H-G-A & 4920 & G-T-A & 5300 \\
\hline
\end{tabular}
\end{center}
\vskip -0.2truecm
\centerline{3 interferometer networks}

\begin{center}
\begin{tabular}{||c|c||c|c||c|c||}
\hline Configuration & $\mathfrak{N}$ & Configuration & $\mathfrak{N}$ & Configuration & $\mathfrak{N}$ \\
\hline V-H-L-G & 5290 & V-H-T-A & 5300 & H-L-G-T & 5340 \\
\hline V-H-L-T & 5150 & V-L-G-T & 5190 & H-L-G-A & 5260 \\
\hline V-H-L-A & 4970 & V-L-G-A & 5080 & H-L-T-A & 4870 \\
\hline V-H-G-T & 5190 & V-L-T-A & 5170 & H-G-T-A & 5220 \\
\hline V-H-G-A & 5140 & V-G-T-A & 5230 & L-G-T-A & 5290 \\
\hline
\end{tabular}
\end{center}
\vskip -0.2truecm
\centerline{4 interferometer networks}

\begin{center}
\begin{tabular}{||c|c||c|c||c|c||}
\hline Configuration & $\mathfrak{N}$ & Configuration & $\mathfrak{N}$ & Configuration & $\mathfrak{N}$ \\
\hline V-H-L-G-T & 5270 & V-H-L-T-A & 5240 & V-L-G-T-A & 5290 \\
\hline V-H-L-G-A & 5230 & V-H-G-T-A & 5330 & H-L-G-T-A & 5290 \\
\hline
\end{tabular}
\end{center}
\vskip -0.2truecm
\centerline{5 interferometer networks}


\begin{references}

\bibitem{geo}
http://www.geo600.uni-hannover.de/

\bibitem{ligo} 
http://www.ligo.caltech.edu/

\bibitem{tama}
http://tamago.mtk.nao.ac.jp/

\bibitem{virgo}
http://www.virgo.infn.it/

\bibitem{aciga}
http://www.anu.edu.au/Physics/ACIGA/

\bibitem{arnaud_burst_1}
N. Arnaud {\it et al.}, \prd~{\bf 59} 082002 (1999).

\bibitem{bala} 
W. G. Anderson and R. Balasubramanian, \prd~{\bf 60} 102001 (1999).

\bibitem{mohanty} 
S. D. Mohanty,\prd~{\bf 61} 122002 (2000).

\bibitem{powermonit}
W. G. Anderson {\itshape et al.}, \prd~{\bf 63} 042003 (2001).

\bibitem{pradier}
T. Pradier {\itshape et al.}, \prd~{\bf 63} 042002 (2001).

\bibitem{vicere}
A. Vicer\'e, \prd~{\bf 66} 062002 (2002).

\bibitem{gursel_tinto} 
Y. G\"ursel and M. Tinto, \prd~{\bf 40} 3884 (1989). 

\bibitem{jk5} 
P. Jaranowski, A. Krolak, \prd~{\bf 49} 1723 (1994).

\bibitem{arnaud_burst_2}
N. Arnaud {\it et al.}, \prd {\bf 65} 042004 (2002).

\bibitem{IGEC}
http:://igec.lnl.infn.it

\bibitem{IGEC_papers}
Z.A. Allen {\it et al.}, \prl~{\bf 85} 5046 (2000); \\
P. Astone {\it et al.}, \cqg~{\bf 18} 243 (2001); \\
P. Astone {\it et al.}, \cqg~{\bf 19} 5449 (2002).

\bibitem{Bose}
S. Bose, S.V. Dhurandhar and A. Pai, {\sl Pramana J. Phys.}~{\bf 53} 1125 (1999). \\
S. Bose, A. Pai and S.V. Dhurandhar, {\sl Int. J. Mod. Phys D}~{\bf 9} 325 (2000).

\bibitem{Finn}
L.S. Finn \prd~{\bf 63} 102001 (2001).

\bibitem{thorne87} K.S. Thorne, {\it Gravitational radiation} in {\bf 300 years of gravitation}, edited by S.W. Hawking and W. Israel (Cambridge University Press, Cambridge, 1987).

\bibitem{PDB} 
A. Pai, S. Dhurandhar, S. Bose, \prd {\bf 64} 042004 (2001).

{\bf
\bibitem{Bose2}
S. Bose, {\it Class. Quantum Grav.} {\bf 19} 1437-1442 (2002).

\bibitem{Bose3}
A. Pai, S. Bose and S. Dhurandhar, {\it Class. Quantum Grav.} {\bf 19} 1477-1483 (2002).
}
\bibitem{Owen}
B.J. Owen, \prd~{\bf 53} 6749 (1996).

\bibitem{Searle}
A. C. Searle, S. M. Scott, D. E. McClelland, {\em Class.Quant.Grav.} {\bf 19} 1465-1470 (2002).

\bibitem{jks} P. Jaranowski, A. Krolak, B.F. Schutz, \prd {\bf 58,} 063001 (1998).

\bibitem{ZM}
T. Zwerger \& E. M\"uller, \aeta~{\bf 320} 209 (1997).

\bibitem{DFM}
H. Dimmelmeier, J.A. Font, and E. M\"uller \aeta~{\bf 388}, 917, (2002); \\
H. Dimmelmeier, J.A. Font, and E. M\"uller \aeta~{\bf 393} 523 (2002).

\bibitem{hello_timing}
N. Arnaud {\it et al.}, \prd~{\bf 67} 062004 (2003).

\bibitem{arnaud_these}
N. Arnaud {\em PhD Thesis} Paris XI Orsay University (2002). \\
Available at www.lal.in2p3.fr/presentation/bibliotheque/publications/Theses02.html

\end{references}
\end{document}